\documentclass[journal,onecolumn,12pt]{IEEEtran}
\usepackage[T1]{fontenc}
\usepackage{blindtext}
\usepackage{graphicx}
\usepackage{setspace}
\usepackage{bm, amsmath, amssymb,amsthm}
\usepackage{mathrsfs}
\usepackage{cite}
\usepackage{bigints}
\usepackage{cases}
\usepackage{subcaption}
\usepackage{color}
\usepackage{multirow}
\usepackage{tablefootnote}
\usepackage{caption}
\usepackage{subcaption}
\usepackage{array}
\usepackage{relsize}
\usepackage{stfloats}
\usepackage{lipsum,multicol}
\usepackage[font={small}]{caption}
\usepackage{caption,amsmath,amssymb,array,amsfonts,float,multicol,cite,xcolor,subcaption,caption,multicol,multirow}
\usepackage{tikz}
\DeclareGraphicsExtensions{.pdf,.png,.jpg}

\DeclareMathOperator{\E}{\mathbb{E}}
\DeclareMathOperator{\C}{\mathbb{C}}
\newtheorem{theorem}{Theorem}[section]

\newtheorem{lemma}[theorem]{Lemma}
\newtheorem{remark}{Remark}

\newcommand{\hsd}{|h_{sd}|^2}
\newcommand{\gsp}{|g_{sp}|^2}

\newcommand{\hsri}{|h^i_{sr}|^2}

\newcommand{\hrHatid}{|h^{\hat{j}}_{rd}|^2}

\newcommand{\grjp}{|g^j_{rp}|^2}
\newcommand{\grHatip}{|g^{\hat{j}}_{rp}|^2}
\newcommand{\lsr}{\lambda_{sr}}
\newcommand{\lsd}{\lambda_{sd}}
\newcommand{\lrd}{\lambda_{rd}}
\newcommand{\lsp}{\lambda_{sp}}
\newcommand{\lrp}{\lambda_{rp}}
\newcommand{\gth}{\gamma_{th}}

\newcommand{\bhsr}{\|{\bm h}_{sr}\|^2}
\newcommand\copyrighttext{ %
	\footnotesize ``This work has been submitted to IEEE for possible publication. Copyright may be transferred without prior notice, after which this version may no longer be accessible."}
\newcommand\copyrightnotice{%
	\begin{tikzpicture}[remember picture,overlay]
	\node[anchor=south,yshift=760pt] at (current page.south) {{\parbox{\dimexpr\textwidth-\fboxsep-\fboxrule\relax}{\copyrighttext}}};
	\end{tikzpicture}%
}
\begin{document}
	\title{Performance of Two-Hop Cooperative Cognitive
	Networks with an Energy Harvesting Relay}
	\author{Komal~Janghel,~\IEEEmembership{Member,~IEEE}, and 
		Shankar~Prakriya,~\IEEEmembership{Senior Member,~IEEE}
		\thanks{\hspace*{-0.3cm}\hrulefill}
		\thanks{K. Janghel is with the Department of Electronics and Communication Engineering, Indian Institute of Information Technology Guwahati, Assam 781015, India (e-mail: komal@iiitg.ac.in).}
		\thanks{S. Prakriya is with the Department of Electrical Engineering, Indian Institute of Technology Delhi, New Delhi 110016, India (e-mail: shankar@ee.iitd.ac.in).}	
	}
	\maketitle
	\vspace{-1.5cm}
	\begin{abstract}
	We consider a simultaneous wireless information and power transfer (SWIPT) enabled energy harvesting (EH)  relay based cooperative cognitive radio network (CCRN), and investigate its outage and throughput performance with the  peak-interference type of power control for both time-switching (TS) and power-splitting (PS) protocols. We assume that the multi-antenna equipped relay uses maximum ratio combining and transmit antenna selection in the first and second hop respectively. Unlike other literature on this topic, we optimally combine  the direct and relayed signals at the destination.  Due to random nature of the channels and the power control used,  considerable performance improvement results. We also analyze the performance of EH-CCRN with incremental relaying, and { analytically quantify the gain in throughput {over} EH-CCRN}. For small number of antennas at the relay, throughput performance of EH-CCRN is actually inferior to one that uses the direct link without  the relay, whereas that of the incremental scheme is always superior.   We establish that there exist optimum values of EH parameters that result in maximum throughput with TS and PS EH protocols. We derive closed-form expressions for these in some special cases. Computer simulations are used to validate the derived expressions.
\end{abstract}
\IEEEpeerreviewmaketitle
\copyrightnotice
\section{Introduction}
There are two major factors limiting growth of wireless communication services. The first relates to acute spectrum scarcity.  Cognitive radio networks (CRNs), because of their ability to increase spectrum utilization efficiency, can alleviate this problem. Of particular interest in this context are underlay CRNs where secondary (unlicensed) users transmit concurrently with those of the (licensed) primary  network (PN), while ensuring  that  performance of the latter is maintained at an acceptable level despite the interference caused.    The second major factor limiting growth of wireless services relates to small battery life-times. Use of energy harvesting (EH) is well motivated to increase  longevity of nodes. Characterizing the performance of CRNs with EH nodes is therefore of great practical interest.
\par    In  \cite{Bhowmick2016,Miridakis2016}, an interweave framework is considered for EH CRN.  Specifically,  \cite{Miridakis2016} proposes a cooperative CRN framework in which one or multiple secondary relays harvest energy and forward source data to secondary destination when  the primary source is sensed to be absent. Performance of the secondary network (SN) is analyzed in terms of energy consumption and outage probability. In \cite{He2016} and \cite{Wang2016}, an overlay approach is used wherein the secondary nodes assist the primary transmission, thereby boosting the signal-to-noise ratio (SNR) at the primary receiver, and enabling secondary transmission.  
 In \cite{Janghel2017},  a joint energy and spectrum sharing  framework {is} suggested wherein the secondary (cognitive) underlay transmitter harvests energy from the primary source using the time-switching (TS) EH protocol before commencing transmission, and its performance is analyzed. In \cite{Mousavifar2014} and \cite{Liu2015},  multiple secondary source and relay nodes  harvest energy  from the primary RF signal using the TS-EH protocol before transmission.In \cite{Yuanwei}, performance of a two-hop secondary network is analyzed assuming that the secondary nodes drawn energy from the primary signal. In \cite{Jia}, a set of relays harvest energy from the primary signal, and from amongst these is picked to relay the secondary information. In both \cite{Yuanwei} and \cite{Jia}, the direct channel from source to destination is assumed to be shadowed, and not used by the destination.  In \cite{LuGe}, a power beacon provides energy to secondary nodes to facilitate multihop relaying. \cite{Janghel2018} analyzes performance of a two-hop with relay powered by secondary source transmission. In  \cite{Janghel2018},  a protocol is suggested that allows energy to be transmitted in packets until the relay has sufficient energy. It is shown to yield good performance with  little to no channel knowledge.  In all these works, the direct link between the source and destination is ignored. 
\par Performance of networks with energy harvesting nodes is limited by the small amount of energy that is harvested.  In \cite{Huang2014,Tran2015,Nguyen2016}, there has been some focus on use of multiple antennas at the relay to harvest energy,  though not in the underlay cognitive radio context. Use of multiple antennas increases the amount of harvested energy, and results in larger throughput since it allows for maximal-ratio combining/ transmission (MRC/MRT) or 	antenna selection. To date however, the advantages of using  multiple antennas at the relay have not been quantified in the underlay EH context.  Another approach to boost performance could be to optimally combine the signals from the direct and the relayed paths.  Cognitive dual hop networks that include the source (S) to destination (D) link have been studied in the past \cite{Tourki2013,Huang2013,Chu2014,Tourki2014}, but mostly in the case when the nodes are self-powered and do not use EH. As noted already, most literature on EH-cooperative CRN (EH-CCRN) has also not considered the direct S-D link in the analysis, which may not always be reasonable since the secondary nodes are located close to each other due to power constraints on the nodes (this is even more so when EH is used). In the non-cognitive context, \cite{Lee2017,Van2016} have studied the influence of the direct path in cooperative communication and analyzed the outage performance.   Recently\footnote{\cite{Xie2018} appeared after presentation of the conference version of this paper \cite{Janghel2017WPMC}.}, \cite{Xie2018} have considered a multiuser network with an energy harvesting relay assisting communication to a destination, assuming the direct path between secondary source and secondary destination. However, they do not use optimal combining, and rely solely on selection combining at D. Neither they do assume multiple antennas at the relay or study performance with incremental relaying.  To the best of our knowledge, there has been no work on EH-CCRN based on the conventional cooperative protocol (suggested in \cite{Laneman2004}) based on optimal combining at the secondary destination. In summary,  the contributions of this paper are as follows\footnote{A part of the analysis limited to incremental relaying with a single antenna relay, and PS-EH was presented in WPMC 2017 \cite{Janghel2017WPMC}.}:
\begin{enumerate}
	\item We present closed-form expression for outage and throughput of an EH-CCRN with multiple antennas at the EH relay assuming optimal combining at the destination, and bring out importance of the direct path in such networks.
	\item We analyze the performance of incremental relaying (IR) in EH-CCRNs, and demonstrate throughput gains.
	\item We show that use of an EH relay makes sense only when there are multiple antennas at R. For small number of antennas at R (assuming MRC at D), throughput of EH-CCRN is actually inferior to the case when only the direct channel (S-D) is used without a relay.  Use of IR on the other hand always results in better performance, irrespective of the number of antennas.
	\item We anlayze PS-EH and TS-EH protocols, and demonstrate that in this context, performance of PS-EH is vastly superior to that of TS-EH.
	\item We quantify the gain  in throughput  with the optimal scheme due to IR, and also with respect to the case when the direct path is ignored.
	\item We derive expressions for the optimum value of EH parameters in EH-CCRNs, including the case when  IR  is used for the number of antennas at the relay is one.
\end{enumerate}   
The proposed framework and  transmission protocols are  discussed in Section~\ref{sec:system_model} and  Section~\ref{sec:transmission_protocol} respectively. SN performance analysis in terms outage and throughput are discussed in Section~\ref{sec:Outage} and \ref{sec:Throughput} respectively. Analytical expressions derived in the paper are validated through simulation results in Section~\ref{sec:simulations}, and conclusions are presented in Section~\ref{sec:conclusion}.
\par	{\em Notations}: $\E_{\C}(\cdot)$ denotes expectation over the condition/conditions $\C$. $X\sim\exp(\lambda)$ implies that $X$ is exponentially distributed with rate parameter $\lambda$, and ${\cal CN}(0,a)$ denotes the circular normal distribution with mean $0$ and variance $a$. $E_n(\cdot)$, represents the generalized exponential integral defined in \cite[5.1.4]{Abramowitz1964}. $F_{X}(x)$ and $f_{X}(x)$ denote the cumulative distributive function (CDF) and probability density function (PDF) of $X$, and $F_{X|Y}(x)$ denotes the CDF of $X$ given $Y$. ${\bf x}^{H}$ denotes the Hermitian of the vector ${\bf x}$, and $\parallel {\bf x}\parallel$ its second norm.
	\section{System Model}\label{sec:system_model}
	S and D possess a single antenna. The EH relay R is equipped with $L$ antennas. The PN consists of the primary transmitter (not depicted in the figure), and the primary receiver P.  Denote the channel between S and the $i^{th}$ antenna of R by $h^i_{sr}\sim {\cal CN}(0,\lambda_{sr}^{-1})$,  and that between the $j^{th}$ antenna of R and D by $h^j_{rd}\sim {\cal CN}(0,\lambda_{rd}^{-1})$, where $i,j=1,2,3,\ldots,L$. Denote the vector channel between S and R by $\bm{h}_{sr}=[h^1_{sr},h^2_{sr},\ldots h^{L}_{sr}]^T$. Similarly,  denote the channel between S and P by   $g_{sp}\sim {\cal CN}(0,\lambda_{sp}^{-1})$, and that between the $j^{th}$ antenna of R and P by $g^j_{rp}\sim {\cal CN}(0,\lambda_{rp}^{-1})$.  We assume a path-loss  model, so that the variance $\lambda_{x}$ of a channel between two nodes is proportional to $d_{x}^{\epsilon}$ where $d_{x}$ is the distance between the nodes, and $\epsilon$ is the path-loss exponent.
	\par In this paper, we make the following assumptions: 
	\begin{enumerate}
		\item As in any other work related to underlay CRNs with peak interference type of power control, we assume that both S and R estimate the channel gains ($\gsp$ and $\grjp$) to P by observing the reverse channel of
		the primary, or by using pilots transmitted by P\cite{Jovicic2009}.
		\item  R does not use its own energy to relay information. It is equipped with a super-capacitor, and acts as an EH node with EH factor $\eta$. It cannot store charge over long intervals, and uses all the energy harvested in the first phase to relay the signal
		to D.
		\item As in most work related to dynamic spectrum access \cite{Tourki2014,Lee2011,Rezki2012,Kashyap2013}, we assume that that the primary signal can be neglected at the secondary receiving nodes (R in the first phase and D in the second). This is reasonable, since the primary nodes are distant from the secondary nodes. Further, the secondary nodes need to be close to each other {\em as compared to their distance to the primary nodes} in order to maintain a reasonable quality of service (QoS). There are two reasons for this. Firstly, the transmit powers in underlay cognitive radio are random, and the degradation in performance due to SNR variation limits the distance between secondary nodes (relaying is still required due to variations in transmit powers and their low average values). Secondly, EH is practical only over short distances. This means that the secondary nodes are close to each other. This assumption on primary interference at the secondary nodes has also been justified on information theoretical grounds \cite{Jovicic2009}.
		\item All the channels are reversible, and quasi-static by nature. We assume that S does not possess knowledge of $\hsd$ and $\bm h_{sr}$.
		\item The multi-antenna relay R employs MRC for EH and reception of signal from S, and transmit antenna selection for relaying the signal to D.  In the second hop (R-D), employing MRT is not practical since  imposing the interference temperature constraints requires knowledge of both magnitude and phase of the channels from each antenna to the primary receiver. Antenna selection is therefore used instead in the second hop.
	\end{enumerate}
	\vspace{-0.15cm}
	\section{Transmission Protocol}\label{sec:transmission_protocol}
	In this paper, we present analysis of performance that is applicable to both TS-EH and PS-EH protocols\cite{Nasir2013}. We briefly review each of these in what follows. 
	\subsection{EH-CCRN with Power-Splitting Protocol}
	\par
\begin{figure}[h]
		\vspace{0cm}
		\centering
		\includegraphics[width=\textwidth]{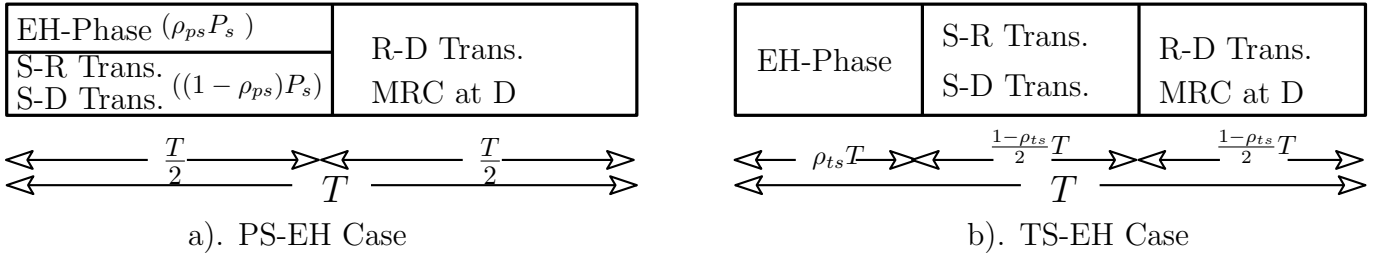}
		\caption{Transmission Scheme}
		\label{fig:trans_scheme}
		\vspace{-0.1cm}
	\end{figure}

 PS-EH is accomplished in two time-slots of duration  $T/2$ as depicted in  Fig.\ref{fig:trans_scheme}a).
	In the first time-slot, S transmits symbols $x$ at rate $R_s$ with power $P_s$ in underlay mode to R and D. Denote the received signal at the $i^{th}$ antenna of the relay R by $y_{r,i-ps}$.  Let ${\bf y}_{r-ps}= [y_{r,1-ps},y_{r,2-ps},\ldots,y_{r,L-ps}]^{T}$.  In PS-EH case, a component of the received signal at R with fraction $\rho_{ps}$ of the received signal power is utilized for EH, while the remaining component ${\bf y}_{r-ps}$  is used to decode the signal. Clearly, ${\bf y}_{r-ps}$ at R and  $y_{d_1}$ at D in the first phase are given by:
	\begin{IEEEeqnarray}{rcl}
	{\bf y}_{r-ps} &=& \sqrt{(1-\rho_{ps}) P_{s}}\,\bm h_{sr}\,x+ {\bf n}_{r}\;\quad\text{and}\label{eq:yr_PS}
	\end{IEEEeqnarray}
\begin{IEEEeqnarray}{rcl}
	y_{d_1} &= &\sqrt{P_{s}}\, h_{sd}\,x+n_{d_{1}}\, \label{eq:yd1_PS}
	\end{IEEEeqnarray}
	respectively, where ${\bf n}_{r}$ is a vector of noise samples with elements $n_{r,i}$, {and} $n_{d_1}$ and $n_{r,i}\sim {\cal CN}\left(0,N_o\right)$. Clearly, the SNR $\Gamma_{d_1}$ at D in the first phase is $P_s\hsd/N_o$.  We assume that R uses a beamformer with weights ${\bf h}_{sr}^{H}/\parallel {\bf h}_{sr}\parallel$. With $\eta$ denoting the EH efficiency, the energy harvested at R is $\eta \rho_{ps}\,P_s \|\bm h_{sr}\|^2 T/2$ (ignoring noise) so that the power $^hP_{r-ps}$  available for the second phase of duration $T/2$ is given by:
	\begin{equation*}
	^hP_{r-ps}=\rho_{ps}\,\eta P_s \bhsr=\beta_{ps} P_s \bhsr\,,\label{eq:Prh-PS}
	\end{equation*}
	\vspace{-0.1cm}
	where $\beta_{ps}= \eta\, \rho_{ps}$, and $\bhsr=\displaystyle\sum_{i=1}^{L}\hsri$.
 Let $I$ denote the interference temperature limit.   In order to ensure that the interference caused to  P is limited to $I$, the power $P_s$ at S  is chosen to be:
	\begin{equation}
	P_s= {I}/{\gsp}\,.\label{eq:PsOnlyInterference}
	\end{equation}
	We consider only the peak interference constraint at S, and ignore the peak power constraint for the following reasons:
	\begin{enumerate}
		\item It is well known that performance of CRNs exhibits an outage floor, and does not improve with increase in peak power (it is in this low outage and high throughput region that CRNs are typically operated) \cite{Tourki2013,Duong2012,Lee2011}. Since the outage with finite peak power is the same as that with infinite peak power, we can simplify the analysis to obtain an estimate of performance of the system in this low outage and high throughput regime, as well as to perform optimizations.
		\item Since performance of CRNs is typically limited by interference, and sufficient peak power is typically available, this assumption is quite reasonable.
	\end{enumerate}   
 Clearly the SNR $\Gamma_{r-ps}$ at the relay for information processing is given by $\Gamma_{r-ps} = (1-\rho_{ps})P_s\parallel {\bf h}_{sr}\parallel^{2/}N_o$. 
	\par In the second time-slot, let  the $j$th antenna at R be used to forward the decoded symbol $\hat{x}$ to D (we discuss antenna selection procedures in Section~\ref{sub:AntennaSelection}). In order to ensure that the interference at P is constrained to $I$, the total transmit power $ P^{j}_{r-ps} $ at R is chosen to be:
	\begin{equation*}
	P^{j}_{r-ps} = \min\left(^hP_{r-ps}, {I}/{|g^j_{rp}|^2} \right). \label{eq:Pr_PS}
	\end{equation*}
	The received signal $y^j_{d_2-ps}$ at D can be expressed as:
	\begin{equation}
	y^j_{d_2-ps} = \sqrt{P^{j}_{r-ps}}\,h^j_{rd}\, \hat{x}+n_{d_2}\;, \label{eq:yd2_PS}
	\end{equation} 
	where $n_{d_2}\sim{\cal CN}(0,N_o)$ is the additive noise.  Clearly, the SNR $\Gamma_{d_{2}-ps}^{j}$ at D in this phase is $\Gamma_{d_{2}-ps}^{j}= P_{r}^{j}|h_{rd}^{j}|^{2}/N_o$. We assume that D uses MRC to combine the signals  (\ref{eq:yd1_PS}) and (\ref{eq:yd2_PS}) obtained (respectively) in the first and second phases. SNRs at R and D can be expressed  as:
	\begin{subequations}\label{eq:SNR_ps}
	\begin{eqnarray}
	\Gamma_{r-ps} &=&\dfrac{(1-\rho_{ps})P_s\bhsr}{N_o}\qquad \text{and} \\
	\Gamma^j_{d-ps} &=& \underbrace{\frac{P^{j}_{r-ps}|h^j_{rd}|^2}{N_o}}_{\Gamma^j_{d_2-ps}}+\underbrace{\frac{P_s\hsd}{N_o}}_{\Gamma_{d_1}}.\label{eq:SNR_dps}
	\end{eqnarray}
	\end{subequations}
\subsection{EH-CCRN with Time-Switching Protocol} \label{sub:EHCCRNwithTSEH}
\par In this case signalling takes place in three time slots as depicted in Fig.\ref{fig:trans_scheme}b). The first slot of duration $\rho_{ts} T$ is used to harvest the energy from the source signal, and remaining  $(1-\rho_{ts})T$ interval is divided into two equal slots to perform S-R and S-D transmissions respectively \cite{Nasir2013}.  Once again, we assume that R uses beamformer weights ${\bf h}_{sr}^{H}/\| {\bf h}_{sr}\|$. Assuming MRC for combining signals at D, SNRs at R and D can be shown to be:
\begin{subequations}\label{eq:SNR_ts}
\begin{eqnarray}
\Gamma_{r-ts} &=&\dfrac{P_s\bhsr}{N_o}\qquad\text{and} \label{eq:SNR_rts} \\
\Gamma^j_{d-ts} &=& \underbrace{\frac{P^{j}_{r-ts}|h^j_{rd}|^2}{N_o}}_{\Gamma^j_{d_2-ts}}+\underbrace{\frac{P_s\hsd}{N_o}}_{\Gamma_{d_1}}.\label{eq:SNR_dts} 
\end{eqnarray}
\end{subequations}
The SNRs of PS-EH and TS-EH from \eqref{eq:SNR_ps} and \eqref{eq:SNR_ts} respectively can be written in a unified form as:
\begin{subequations}
\begin{IEEEeqnarray}{rcl}
	\Gamma_r &=&\dfrac{\xi\,P_s\bhsr}{N_o}\quad\text{and} \ ^hP_r=\beta P_s  \parallel {\bf h}_{sr}\parallel^{2} \label{SNR_R} \\
	\Gamma^{j}_d &=&  \underbrace{\dfrac{1}{N_o}\overbrace{\min\left(^hP_r, \dfrac{I}{|g^j_{rp}|^2} \right)}^{P_r^j}\,|h^j_{rd}|^2}_{\Gamma_{d_2}^{j}}+ \underbrace{\dfrac{P_s\,|h_{sd}|^2}{N_o}}_{\Gamma_{d_1}}\, , \label{SNR_D}
\end{IEEEeqnarray}
\end{subequations}
where $\xi$ and $\beta$ take values as listed in Table-\ref{tab:commparms} for PS-EH and TS-EH protocols.
\begin{table}[!htb]
	\centering
	\renewcommand{\arraystretch}{1.2}
	\begin{tabular}{ |c|c|c|c| } 
		\hline
		S.N. & Parameter & PS-EH& TS-EH  \\ \hline
		1. &$\rho$  &$\rho_{ps}$ &$\rho_{ts}$\\\hline
		2. & $\xi$   & $(1-\rho_{ps})$& 1 \\ \hline
		3. & $\beta$ & $\beta_{ps}=\eta \rho_{\rho_{ps}}$& $\beta_{ts}=\frac{2\eta\,\rho_{ts}}{(1-\rho_{ts})}$ \\ \hline
	\end{tabular}
	\caption{Parameters for PS-EH and TS-EH.\label{tab:commparms}}
\end{table}
\subsection{EH-CCRN with Incremental Relaying} \label{sub:EHCCRNwithIR}
\begin{figure}[h]
\centering
		\includegraphics[width=0.88\textwidth]{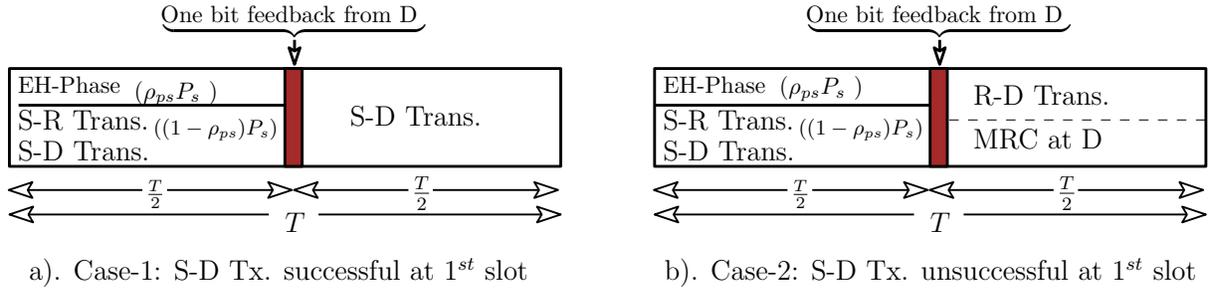}
		\caption{Transmission scheme for IR with PS-EH protocol}
		\label{fig:transmission_schemeIncPS}
		\vspace{-0.5cm}
\end{figure}

The transmission scheme for the PS EH-CCRN with IR protocol is depicted in Fig.~\ref{fig:transmission_schemeIncPS}. In the first transmission time-slot, S transmits information symbols to D and R. R harvests energy from this signal using power splitting, and attempts to decode the information symbols. Meanwhile,  D which has SNR  $P_s\hsd/N_o$ attempts to decode the symbols.   If successful, it sends a one bit feedback to S and R\footnote{We assume that feedback time is extremely small and can be neglected in the analysis without loss of generality.}. R then discards the decoded symbols, and S transmits a {\em new} block of symbols to D as depicted in Fig.\ref{fig:transmission_schemeIncPS}a. Else, R relays the symbols using the energy harvested, and D combines the signals in the first and the second time-slots as depicted  in Fig.\ref{fig:transmission_schemeIncPS}b.
\begin{figure}[h]	
	\centering	
		\includegraphics[width=0.88\textwidth]{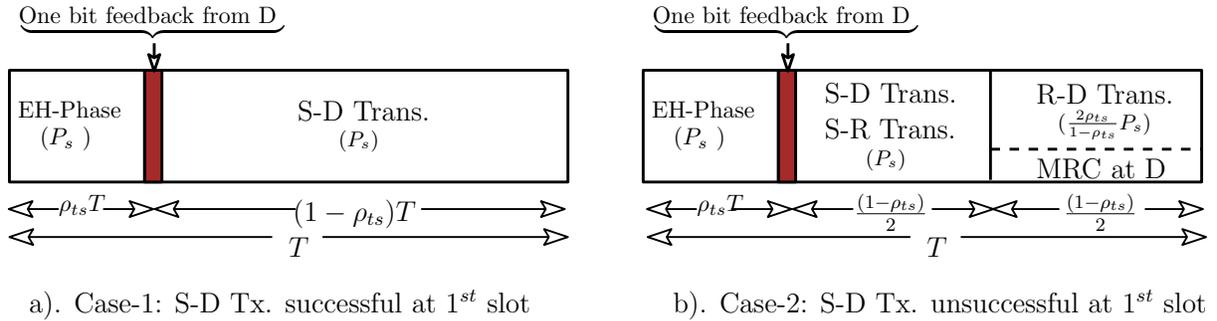}
		\par\caption{Transmission scheme for IR with TS-EH protocol}
		\label{fig:transmission_schemeIncTS}
\end{figure} 

\par In the TS-EH-CCRN with IR, the first time-slot of duration $\rho_{ts} T$ is reserved for EH. D attempts to decode these energy symbols, and send a one bit feedback to S and R.  If D send '1' as feedback,  rest of the signalling duration is used for S-D transmission as shown in  Fig.\ref{fig:transmission_schemeIncTS}a.  If D send a feedback of '0', then a two-hop signalling is used as shown in Fig.\ref{fig:transmission_schemeIncTS}b.
\subsection{Antenna Selection at Relay (R)} \label{sub:AntennaSelection}
As noted already, R performs MRC to maximize SNR in the S-R link. For the second hop, it uses transmit antenna selection (TAS) to to maximize the SNR  at D from (\ref{SNR_D}) as follows:
\begin{eqnarray}
	\hat{j}&=& \arg\underset{j}{\max}\left(\Gamma^j_{d}\right)=\arg\underset{j}{\max}\left(\Gamma^j_{d_2}\right)=\arg\underset{j}{\max}\left(\min\left(^hP_r, {I}/{|g^j_{rp}|^2}\right)|h^j_{rd}|^2\right)\nonumber.
	\end{eqnarray}
\par In subsequent sections, performance analysis is presented in terms of outage probability and throughput  assuming fixed-rate signalling at rate $R_s$.
\section{Outage Analysis}\label{sec:Outage}
In this section, we analyze performance of the SN in terms of outage probability assuming MRC is used at D. We also analyze the outage probability when IR protocol is used.
\subsection{Outage Probability of EH-CCRN with S-D link}
From the SNRs (\ref{SNR_R}) and (\ref{SNR_D}), the outage probability with the DF relay is given by \cite[eq.17]{Laneman2004}:
\begin{IEEEeqnarray}{rcl}
p &=&\Pr\left[\min\left(\Gamma_r,\Gamma^{\hat{j}}_d \right)<\gamma_{th}\right]=\underbrace{\Pr\left[\Gamma_r< \gamma_{th}\right]}_{p_1} + \underbrace{\Pr\left[\Gamma^{\hat{j}}_d< \gamma_{th},\;\Gamma_r \geq \gamma_{th}\right]}_{p_2},\label{eq:pout}
\end{IEEEeqnarray}
where $\gamma_{th}=2^{2 R_s}-1$, and $R_s$ is the target rate of the SN. We first evaluate $p_1$ as follows:
\begin{eqnarray}
p_1=\Pr\left[\Gamma_r< \gamma_{th}\right] = \Pr\left[\frac{\xi P_s\bhsr}{N_o}< \gamma_{th}\right]. \label{eq:p1}
\end{eqnarray}
We know that the $\bhsr$ is Gamma distributed and its PDF is given by:
\begin{eqnarray}
f_{\bhsr}(x)= \frac{x^{L-1} \lsr^L}{(L-1)!}e^{-\lsr \,x}. \label{eq:pdf_bhsr}
\end{eqnarray}
 Using $P_s$ from (\ref{eq:PsOnlyInterference}), $p_1$ can be easily shown to be:
{\begin{equation}
p_1=\Pr\left[\frac{\bhsr}{|g_{sp}|^2}< \frac{\psi}{\xi}\right]=\left(\frac{\lsp \xi}{\lsr \psi}+1\right)^{-L}\,,\label{eq:p1Final}
\end{equation}}
where $\psi =\frac{\gth}{I/N_o}$.
Using (\ref{SNR_R}) and (\ref{SNR_D}),  $p_2$ can be expressed as:
\begin{IEEEeqnarray}{rcl}
p_2&=&\Pr\Bigg[\frac{P_s \hsd}{N_o}+\frac{1}{N_o}{\displaystyle\max_{j}}\left(\min\left(^hP_r, \frac{I}{|g^j_{rp}|^2}\right)|h^j_{rd}|^2\right) < \gamma_{th},\frac{\xi P_s \bhsr}{N_o} \geq \gamma_{th}\Bigg].\label{eq:p2_def}
\end{IEEEeqnarray}
Unfortunately, it is not possible to derive an exact expression for $p_2$. Instead, we show in Appendix-A that $p_2$ can be approximated as follows:
\begin{IEEEeqnarray}{rcl}
		p_2&\approx&\sum _{i=0}^L \binom{L}{i} \sum _{j=0}^i (-1)^{i+j} t^j \binom{i}{j} \Bigg(\frac{\lsr^L \lrd^2 \lsp i^2}{(L-1)!\beta^2 \lsd^2 \left(\frac{\lrd \lsp i}{\beta \lsd}+\frac{\lrd i \psi}{\beta}+\frac{\lrp j}{\beta}\right)}\Bigg(\sum _{k=0}^{L-2} \Gamma (-k+L-1) \left(\frac{\lrd i}{\beta \lsd}\right)^k \nonumber\\
		&&\times\bigg(e^{-\frac{\lrp j \xi}{\beta \psi}} \left(\frac{\xi (\lsd \psi+\lsp)}{\psi}+\lsr\right)^{k-L+1}-\left(\frac{\lsp \xi}{\psi}+\lsr\right)^{k-L+1} e^{-\frac{\xi \left(\frac{\lrd i \psi}{\beta}+\frac{\lrp j}{\beta}\right)}{\psi}}\bigg)+\left(\frac{\lrd i}{\beta \lsd}\right)^{L-1} \nonumber\\
		&&\times\left\{\text{Ei}\left(\frac{\lrd i}{\beta \lsd} \left(\lsr+\frac{\lsp \xi}{\psi}\right)\right)-\text{Ei}\left(\frac{\lrd i }{\beta \lsd}\left(\lsr+\lsd \xi+\frac{\lsp \xi}{\psi}\right)\right)\right\}\exp \left(-\frac{\xi}{\psi} \left(\frac{\lrd i \psi}{\beta}+\frac{\lrp j}{\beta}\right)\right)\nonumber\\
		&&\times \exp \left(-\frac{\lrd i}{\beta \lsd} \left(\frac{\lsp \xi}{\psi}+\lsr\right)\right)\Bigg)+\frac{\lrp^2 \lsp \lsr j^2 e^{\frac{\lrp \lsr j}{\beta (\lsd \psi+\lsp)}} }{\beta (\lsd \psi+\lsp)^2 \left(\frac{\lrd \lsp i}{\lsd}+{\lrd i \psi}+{\lrp j}\right)}\nonumber\\
		&&\times \Bigg\{E_L\left(\frac{\lrp \lsr j}{\beta (\lsp+\lsd \psi)}\right)-\left(\frac{1}{\frac{\xi (\lsd \psi+\lsp)}{\psi\lsr}+1}\right)^{L-1} E_L\left(\frac{{\lrd i \psi}+{\lrp j}}{\beta\lsp} \left(\lsr+\frac{(\lsp+\lsd \psi) \xi}{\psi}\right)\right)\Bigg\}\nonumber\\
		&& -\frac{\lsr \left({\lrd i \psi}+{\lrp j}\right)^2 e^{\frac{\lsr}{\lsp} \left(\frac{\lrd i \psi}{\beta}+\frac{\lrp j}{\beta}\right)}}{\beta\lsp \left(\frac{\lrd \lsp i}{ \lsd}+{\lrd i \psi}+{\lrp j}\right)}\Bigg\{E_L\left(\frac{\lsr}{\lsp} \left(\frac{\lrd i \psi}{\beta}+\frac{\lrp j}{\beta}\right)\right)-\left(\frac{\lsp \xi}{\psi\lsr}+1\right)^{1-L}  \nonumber\\
		&&\times E_L\left(\frac{\left({\lrd i \psi}+{\lrp j}\right)}{\beta\lsp} \left(\lsr+\frac{\lsp \xi}{\psi}\right)\right)\Bigg\}-\left(\frac{\lsr}{\frac{\lsp \xi}{\psi}+\lsr}\right)^L e^{-\frac{1-\rho}{\psi} \left(\frac{\lrd i \psi}{\beta}+\frac{\lrp j}{\beta}\right)}\nonumber\\
		&&+\frac{\lsp}{\lsd \psi+\lsp} e^{-\frac{\lrp j \xi}{\beta \psi}} \left(\frac{1}{\frac{\xi (\lsd \psi+\lsp)}{\lsr \psi}+1}\right)^L+\frac{\lsd \psi}{\lsd \psi+\lsp}\Bigg).\label{eq:p2_FinalMRC}
\end{IEEEeqnarray}
where $t$ is given by:
\begin{eqnarray}
	t&=&1-\bigg(1+\frac{\lambda_{rd}}{\lambda_{rp}}\bigg(\gamma_{th} -\frac{ \lambda_{sp}}{\lambda_{sd}} \bigg(\log \bigg(\frac{\gamma_{th} \lambda_{sd} }{ \lambda_{sp}}+1\bigg)+\frac{{\gamma_{th} \lambda_{sd} }}{{\gamma_{th} \lambda_{sd} }+{ \lambda_{sp}}}\bigg)\bigg)\bigg)^{-1}.\label{eq:t}
\end{eqnarray}
As noted already, in underlay CRNs with EH relays, the secondary nodes are close to each other, and distant from P ($g_{rp}^{j}$ has low variance or $\lambda_{rp}$ is very large) for acceptable secondary QoS. Further, the energy harvested  is small (so that $^hP_r$ is small) in practice. For this reason, $ P^{j}_{r} = \min\left(^hP_{r}, {I}/{|g^j_{rp}|^2} \right)$ equals $^hP_{r}$ with a very high probability. In Section~\ref{sec:simulations}, we demonstrate through simulations that the outage and throughput (defined in the Section~\ref{sec:Throughput}) with $P^{j}_{r}$ is indistinguishable from that obtained by replacing it by $^hP_{r}$. It is emphasized that in practical implementations R will continue to impose the peak interference constraint, and the approximation is only used to obtain a simpler expression. We can achieve this approximation simply by using $\lrp\rightarrow\infty$ in \eqref{eq:p1Final},\eqref{eq:p2_FinalMRC} and \eqref{eq:t}. It is clear from \eqref{eq:t} that for $\lrp\rightarrow\infty$, $t\rightarrow0$ so that only one summation term corresponding to $j=0$ is retained in the {second} summation of \eqref{eq:p2_FinalMRC}. Also,  $\sum_{i=0}^{L}\binom{L}{i}(-1)^{i}=0\,\, \text{for } L\neq0$ \cite[eq. 4.2.1.3]{Prudnikov1992}.
Using these facts, the approximated expression $\tilde{p}_2$ of $p_2$ is given as follows:
{\small\begin{IEEEeqnarray}{rcl}
\tilde{p}_2&\approx&\sum _{i=1}^L (-1)^i \binom{L}{i} \Bigg(\frac{\lrd^2 \lsp \lsr^L }{\beta \lsd^2 (L-1)! \left(\frac{\lrd \lsp}{ \lsd}+{\lrd \psi}\right)}\bigg(\sum _{k=0}^{L-2} i\, \Gamma (L-k-1) \left(\frac{\lrd i}{\beta \lsd}\right)^k \bigg(\left(\frac{\xi (\lsd \psi+\lsp)}{\psi}+\lsr\right)^{k-L+1}\nonumber\\
&& -e^{-\frac{\lrd i \xi}{\beta}}\left(\frac{\lsp \xi}{\psi}+\lsr\right)^{k-L+1}\bigg)+\left(\frac{\lrd i}{\beta \lsd}\right)^L \exp \left(-\frac{\lrd\,i}{\beta \lsd} \left(\frac{\lsp \xi}{\psi}+\lsr\right)-\frac{\lrd i \xi}{\beta}\right)\bigg\{\text{Ei}\left(\frac{\lrd i}{\beta \lsd} \left(\lsr+\frac{\lsp \xi}{\psi}\right)\right) \nonumber\\
&&-\text{Ei}\left(\frac{\lrd i}{\beta \lsd} \left(\lsr+\lsd \xi+\frac{\lsp \xi}{\psi}\right)\right)\bigg\}\bigg)-\frac{\left(\lrd^2 \lsr i \psi^2 \right)}{\beta \lsp \left(\frac{\lrd \lsp}{ \lsd}+{\lrd \psi}\right)}e^{\frac{\lrd \lsr i \psi}{\beta \lsp}} \bigg\{E_L\left(\frac{\lrd \lsr i \psi}{\beta \lsp}\right)-\left({\frac{\lsp \xi}{\psi \lsr}+\lsr}\right)^{1-L}\nonumber\\
&&\times E_L\left(\frac{\lrd i \psi}{\beta \lsp} \left(\lsr+\frac{\lsp \xi}{\psi}\right)\right)\bigg\}-e^{-\frac{\lrd i \xi}{\beta}} \left(\frac{1}{\frac{\lsp \xi}{\psi\lsr}+1}\right)^L\Bigg)-\left(\frac{1}{\frac{\lsp (1-\rho)}{\lsr \psi}+1}\right)^L.\label{eq:p2_WtRP}
\end{IEEEeqnarray}}
With this, the approximate outage $\tilde{p}=p_1 + \tilde{p}_2$. While the approximate expression $\tilde{p}$ so obtained is accurate, it is still too complicated to derive useful insights.
\subsubsection{{Further} approximation of overall outage $p$}
{In typical operating conditions, the following relation holds ${\lsp I/N_o}\gg \gth$.
We use this to obtain expressions for throughput optimal $\rho$ in some special cases.}
\begin{lemma}
	When $\left(\frac{\lsp}{\psi} = \frac{\lsp I/N_o}{\gamma_{th}}\gg1\right)$, the overall outage $p$ can be approximated as $\tilde{p}$ where:
	\begin{IEEEeqnarray}{rcl}
	\tilde{p}&\approx&\sum _{i=1}^L (-1)^{i+1} \binom{L}{i} \Bigg(\frac{1 }{ \frac{\lsp}{ \psi \lsd}+1}\frac{i \lrd \lsr \psi}{\beta \lsp}e^{\frac{i \lrd \lsr \psi}{\beta \lsp}} E_L\left(\frac{\lrd \lsr i \psi}{\beta \lsp}\right)+{\left(\frac{\lsp \xi}{\lsr \psi}+1\right)^{-L}}e^{-\frac{i \lrd \xi}{\beta}}\Bigg).\nonumber
	\end{IEEEeqnarray}
\end{lemma}
\begin{IEEEproof}
Proof is presented in Appendix-B.
\end{IEEEproof}	
$\tilde{p}$ above can also be represented in an alternative form by using  $n e^z E_{n+1}(z)+z e^z E_{n}(z)=1$ \cite[5.1.14]{Abramowitz1964}  as follows:
{\begin{IEEEeqnarray}{rcl}
		\tilde{p}	&\approx&\underbrace{\overbrace{\left(1+\frac{\lsp}{\lsd \psi}\right)^{-1}}^{T_{11}}\overbrace{\Big(1-  \sum _{i=1}^L (-1)^{i+1}\binom{L}{i} L e^{\frac{\psi i \lrd \lsr}{\beta \lsp}} E_{L+1}\Big(\frac{\psi i \lrd \lsr}{\beta \lsp}\Big)\Big)}^{T_{12}}}_{T_1}+\underbrace{ \frac{1-\big(1-e^{-\frac{\lrd \xi}{\beta}}\big)^L}{\left({\lsp \xi}/({\lsr \psi})+1\right)^L}}_{T_2}. \label{eq:pHSNR}
\end{IEEEeqnarray}}
\par The above approximations will be used to derive expressions for the throughput-optimum EH parameters ($\rho^*_{ps}$ and $\rho^*_{ts}$) in Section-\ref{sub:TauMRC}.
\subsection{Outage Analysis of EH-CCRN without S-D link}
\par In some situations, the direct channel (S-D) is shadowed or in deep fade. In this case, the destination is  dependent only on the relayed signal. It is this special case that has been studied in EH CRNs \cite{Janghel2018}. Even in non-cognitive cooperative two-hop EH networks, the importance of the direct channel has not been brought out in literature so far. We show in this paper that utilizing the direct channel results in large improvement in throughput in both underlay cognitive and cooperative two-hop networks, especially when IR is used. The outage is once again given by (\ref{eq:pout}), with $\Gamma_r$ given by (\ref{SNR_R}), and $\Gamma_{d}^{\hat{j}}$ modified to: 
\begin{eqnarray}
\Gamma^{\hat{j}}_d &=&  \dfrac{1}{N_o}\min\left(^hP_r, {I}/{\grHatip} \right)\,\hrHatid. \label{SNR_D_nodirect}
\end{eqnarray} 
The outage probability $p_{nd}$ can be obtained from that derived earlier simply by using $\lambda_{sd}\rightarrow \infty$.  Clearly,  $p_{nd} = p_1 + p_{2-nd}$, where $p_{2-nd} = \lim\limits_{\lsd\rightarrow\infty} p_2$.
Some straightforward manipulations on \eqref{eq:p2_FinalMRC}  yields:
{\small\begin{IEEEeqnarray}{rcl}
		p_{2-nd}&\approx&\sum _{i=0}^L \binom{L}{i} \sum _{j=0}^i (-1)^{i+j} t_{nd}^j \binom{i}{j} \Bigg(1-\frac{\lsr \left({\lrd i \psi}+{\lrp j}\right)^2 e^{\frac{\lsr}{\lsp} \left(\frac{\lrd i \psi}{\beta}+\frac{\lrp j}{\beta}\right)}}{\beta\lsp \left({\lrd i \psi}+{\lrp j}\right)} \Bigg\{E_L\left(\frac{\lsr}{\lsp} \left(\frac{\lrd i \psi}{\beta}+\frac{\lrp j}{\beta}\right)\right)\nonumber\\
		&&-\left(\frac{\lsp \xi}{\psi\lsr}+1\right)^{1-L} E_L\left(\frac{\left({\lrd i \psi}+{\lrp j}\right)}{\beta\lsp} \left(\lsr+\frac{\lsp \xi}{\psi}\right)\right)\Bigg\}-\left(\frac{\lsr}{{\lsp \xi}/{\psi}+\lsr}\right)^L e^{-\frac{1-\rho}{\psi} \left(\frac{\lrd i \psi}{\beta}+\frac{\lrp j}{\beta}\right)}\Bigg).\label{eq:p2_NoSD}
\end{IEEEeqnarray}}	
where $t_{nd} = 1-\left(1+{\lambda_{rd} \gamma_{th} N_o}/{\lambda_{rp}}\right)^{-1}$.
In Section~\ref{sec:simulations}, we use the above to demonstrate importance of taking the direct link into consideration.
\section{Throughput Analysis} \label{sec:Throughput}
In this section, we analyze the throughput performance of a cooperative cognitive EH relaying network. We show how the EH parameter can be chosen optimally. We demonstrate that incremental relaying  results in large throughput gains, and also analyze the throughput performance without the direct {link} to bring out this fact.
\vspace{-0.4cm}
\subsection{Throughput Analysis of EH-CCRN with S-D link} \label{sub:TauMRC}
We define the throughput $\tau$ as:
\begin{eqnarray}
\tau&=& 0.5{R_s}\zeta({1-p})\label{eq:tau_mrc},
\end{eqnarray}
where $\zeta$ is $1$ for PS-EH and $(1-\rho)$ for TS-EH.
 $\frac{\zeta}{2}$ represents the effective time available for  information transmission between nodes S and D. We use $\tau$ without any subscript to refer to throughput of the EH-CCRN with S-D link, and use $\tau_{in}$ and  $\tau_{nd}$ to refer to throughputs with incremental relaying and relayed networks without the direct link respectively. When we need to refer to quantities specific to PS-EH and TS-EH, we use $\tau_{ps}$ and $\tau_{ts}$ respectively. 
\subsubsection{ Value of $\rho$ for optimum throughput}
As noted already, optimizing $\rho$ is crucial for attaining good throughput performance. To determine this optimum $\rho$ value, we use the approximate expression for outage $\tilde{p}$ in (\ref{eq:pHSNR}). In this case, $\tau$ can be approximated as $\tilde{\tau}$, and written as:
 $\tilde{\tau}= \frac{R_s}{2}\zeta({1-\tilde{p}})$.
Using \eqref{eq:pHSNR}, we can represent $\tilde{\tau}$ as follows: 
	{\begin{equation}
		\tilde{\tau}\approx \frac{R_s\zeta}{2}\Bigg(\hspace{-0.1cm}1+\left(1+\frac{\lsp}{\lsd \psi}\right)^{-1}\hspace{-0.15cm}\underbrace{\Big(\hspace{-0.1cm}\sum _{i=1}^L (-1)^{i+1}\binom{L}{i} L e^{\frac{\psi i \lrd \lsr }{\beta \lsp}} E_{L+1}\Big(\hspace{-0.1cm}\frac{\psi i \lrd \lsr }{\beta \lsp}\hspace{-0.1cm}\Big)-1\hspace{-0.1cm}\Big)}_{F_1(\rho)}+\underbrace{ \frac{\big(1-e^{-\frac{\lrd \xi}{\beta}}\big)^L-1}{\left(1+{\lsp \xi}/({\lsr \psi})\right)^L}}_{F_2(\rho)}\hspace{-0.05cm}\Bigg)\,. \label{eq:tauMRCLRPtendInfty}
		\end{equation}}
\begin{lemma}\label{lemma:TauConcaveWRTRho}
Throughput of the EH-CCRN is a concave function of $\rho$.
\end{lemma}
\begin{IEEEproof}
	 Proof is based on the fact that individually $F_1(\rho)$ and $F_2(\rho)$ are concave functions of $\rho$, and the fact that the summation of two concave functions is always concave \cite{Boyd2004}.	Please refer to Appendix-C for proof of concavity of functions $F_1(\rho)$ and $F_2(\rho)$ with respect to (w.r.t.) $\rho$. 
\end{IEEEproof}
	\par The optimum value of $\rho$ that maximizes throughput can be derived by solving $\frac{d\tilde{\tau}}{d\rho}=0$.
Unfortunately, it is difficult to obtain a closed-form expression for optimum $\rho$ due to the complex expressions involved. Numerical techniques need to be used. 
 However, for the special case when $L=1$, a closed-form expression is possible. Denote by $\rho^*_{ps}$ and $\rho^*_{ts}$, the optimum $\rho$ for the cooperative cognitive networks employing the PS-EH and TS-EH protocols respectively.
\begin{lemma}
When $L=1$, the value of $\rho$ that maximizes throughput in PS-EH and TS-EH cases can be expressed as:
\begin{subequations}\label{eq:rho_opt}
\begin{IEEEeqnarray}{rcl}
&&\hspace{-0.5cm}\rho^*_{ps} \approx \dfrac{1-\sqrt{\left(1+{\lsp}/({\lsd\psi})\right)}\,{ \psi \lsr}/{ \lsp}}{1+\sqrt{\left(1+{\lsp}/({\lsd\psi})\right)}\,{ \eta}/{\lrd}}\label{eq:rhoPS}\quad \text{and}\\
&&\hspace{-0.5cm}\rho^*_{ts}\approx\dfrac{\frac{2 \eta}{\lrd \lsr \psi} \sqrt{\dfrac{\lsd \lsp \psi}{(\lsd+\lsp/\psi)2\eta/(\lsr\lrd)-1}}-1}{{2 \eta \lsp}/({\lrd \lsr \psi})-1}.\label{eq:rhoTS}
\end{IEEEeqnarray}
\end{subequations}
\end{lemma} 
\begin{IEEEproof}
Derivation is presented in Appendix-D. 
\end{IEEEproof}
 In Section~\ref{sec:simulations}, we demonstrate that the $\rho_{ps}^{*}$ and $\rho_{ts}^{*}$ estimates of (\ref{eq:rhoPS}) and (\ref{eq:rhoTS}) are accurate. 
\subsection{Throughput of EH-CCRN Without S-D Link}
For comparison purpose, throughput in the special case when the direct S-D channel is ignored (or too weak to be used) is of interest, since this is the case that has been considered in most literature so far. In this section, we derive an expression for this throughput and use it for benchmarking the schemes presented in this paper that utilize the direct S-D channel. 
Throughput $\tau_{nd}$ in this case becomes:
\begin{IEEEeqnarray}{rcl}
	\tau_{nd} = 0.5 R_s \zeta (1-p_{nd})\,,\label{eq:tauWithdSDTendToInfty}
\end{IEEEeqnarray}
where $p_{nd} \triangleq p_1+p_{2-nd}$. From \eqref{eq:p1} and \eqref{eq:p2_NoSD}, $\tau_{nd}$ can be evaluated from \eqref{eq:tauWithdSDTendToInfty}. 
\par The throughput $\tilde{\tau}_{nd}$  at {in typical operating conditions (i.e.$\lsp\gg \psi$)}  can be derived directly from \eqref{eq:tauMRCLRPtendInfty} with $\lsd\rightarrow\infty$ and can represented below:
{\begin{equation}
\tilde{\tau}_{nd}\approx						\frac{R_s\zeta}{2}\Bigg(1-\underbrace{\Big(1-  \sum _{i=1}^L (-1)^{i+1}\binom{L}{i} L e^{\frac{\psi i \lrd \lsr m }{\beta \lsp}} E_{L+1}\Big(\frac{\psi i \lrd \lsr m }{\beta \lsp}\Big)\Big)}_{F_1(\rho)}-\underbrace{ \frac{1-\big(1-e^{-\frac{\lrd \xi}{\beta}}\big)^L}{\left(1+{\lsp \xi}/({\lsr \psi})\right)^L}}_{F_2(\rho)}\Bigg)\,.  \label{eq:tauMRCLRPtendInftyWtLSD}
\end{equation}}
 As in Section~\ref{sub:TauMRC}, it can be shown that $\tilde{\tau}_{nd}$ \eqref{eq:tauMRCLRPtendInftyWtLSD} is a concave function $\rho$. Again, there is an optimum value of $\rho$ as with EH-CCRN with S-D link.  We denote by $\rho_{nd-ps}^{*}$ and $\rho_{nd-ts}^{*}$ the optimum $\rho$ with PS-EH and TS-EH protocols when the direct S-D channel is too weak and cannot be used.
\begin{remark}
	In this case for $L=1$ and in the absence of the direct channel, \eqref{eq:rhoPS} and \eqref{eq:rhoTS} can be expressed as:
	\begin{subequations}\label{eq:rho_opt_withoutDSD}
	\begin{eqnarray}
		\rho^*_{nd-ps} &\approx& \dfrac{1-{ \psi \lsr}/{ \lsp}}{1+{ \eta}/{\lrd}}\label{eq:rhoPSdSDInfty}\qquad\quad\text{and}\\
		\rho^*_{nd-ts} &\approx&\dfrac{\sqrt{{2 \eta  \lsp}/({\psi \lrd \lsr })}-1}{{2 \eta  \lsp}/({\psi \lrd \lsr })-1}. \label{eq:rhoTSdSDInfty}
	\end{eqnarray} 
	\end{subequations}
\end{remark}
\begin{lemma}
	The gain in throughput due to use of the direct S-D channel quantified by $\tau-\tau_{nd}$ is finite and positive, and is given by:
	\begin{equation}
	\tau-\tau_{nd}\approx \frac{R_s\zeta}{2}\bigg(\underbrace{1-\sum _{i=1}^L (-1)^{i+1} \binom{L}{i}L \,e^{\frac{\psi i \lrd \lsr }{\beta  \lsp}} E_{L+1}\Big(\frac{\psi i \lrd \lsr}{\beta  \lsp}\Big)}_{F_1(\rho)}\bigg) \left(\frac{\lsp/(\lsd \psi )}{1+\lsp/(\lsd \psi )}\right).\label{eq:TauDiff}
	\end{equation}
	Further, when the number of antennas at R becomes large ($L\rightarrow\infty$), $\tau-\tau_{nd}\rightarrow0$.
\end{lemma}
\begin{IEEEproof} We present an outline of the proof here.
	\eqref{eq:TauDiff} follows from \eqref{eq:tauMRCLRPtendInfty} and \eqref{eq:tauMRCLRPtendInftyWtLSD}, and some simple manipulations. {It can be shown that} $F_1(\rho)=\Pr\left[\frac{\beta I\bhsr\hrHatid}{N_o\gsp}<\gth\right]$. {Since it is a} probability term, its value is less than or equal to $1$, which ensures that $\tau-\tau_{nd}$ is a positive quantity. When $L\rightarrow\infty$, it is shown in  Appendix-E that the following relation holds:  
 \begin{equation}
\sum _{i=1}^L (-1)^{i+1} \binom{L}{i}L \,e^{\frac{\psi i \lrd \lsr }{\beta  \lsp}} E_{L+1}\Big(\frac{\psi i \lrd \lsr}{\beta  \lsp}\Big)\approx 1.\label{eq:SumSum_ijxExpijxE1ijx}
 \end{equation}
 Using the above, it can be shown that  $\tau-\tau_{nd}\rightarrow0$ when $L\rightarrow \infty$. 
 \end{IEEEproof}
\par When $L$ is large, the contribution from the direct path becomes negligible i.e. the relayed path is almost always available for end-to-end transmission. However, for practical small values of $L$, the direct channel contributes significantly to the throughput. Intuitively, the reason is as follows. In CRNs, especially in network with EH relays, the nodes need to close to each other for acceptable QoS. This is primarily because of random nature of the power at S and R  due to the peak interference constraint, and the random nature of the energy harvested. Clearly, SNR of the direct channel is not always insignificant, and it should therefore not to be ignored for practical $L$ values.
\subsection{Throughput of EH-CCRN with IR Protocol}
In the previous subsection, performance of two-hop EH relaying with conventional CCRN was analyzed. However, as is well known, it is not necessarily throughput efficient since it requires two hops for signaling. IR protocol discussed in Section~\ref{sub:EHCCRNwithIR} exploits the fact that when the direct S-D link is strong so that  $\log_{2}\left(1+\Gamma_{d_{1}}\right)=\log_2\left(1+\frac{P_s|h_{sd}|^{2}}{N_o}\right)\geq R_s$,  signaling can be accomplished in just one hop, thereby increasing throughput. 
In this case, throughput (denoted by $\tau_{in}$ with subscript $in$ denoting incremental) can be expressed as:
\begin{IEEEeqnarray}{rcl}
\tau_{in}&=& \zeta\bigg(0.5\,R_s\underbrace{\Pr\left[\Gamma_{d_1}<\gamma_{th},\, \Gamma_{r}\geq\gamma_{th},\, \Gamma^{\hat{j}}_{d}\geq\gamma_{th}\right]}_{q_1}+ R_s\underbrace{\Pr[\Gamma_{d_1}\geq\gamma_{th}]}_{q_2}\bigg)=\zeta (0.5 R_s q_1 + R_s q_2).\label{eq:tauIn}
\end{IEEEeqnarray}
\par To facilitate quantification of the gains due to incremental relaying explicitly,  we write down an expression for throughput of the incremental scheme in terms of that of the EH-CCRN scheme.  To facilitate this, we rewrite $q_1$ in the above equation. Since $\Pr\big[\Gamma_{d_1}<\gamma_{th},\, \Gamma_{r}\geq\gamma_{th},\, \Gamma^{\hat{j}}_{d}\geq\gamma_{th}\big]$$+ \Pr\big[\Gamma_{d_1}\geq\gamma_{th},\, \Gamma_{r}\geq\gamma_{th},\, \Gamma^{\hat{j}}_{d}\geq\gamma_{th}\big]$ $=\Pr\big[\Gamma_{r}\geq\gamma_{th},\, \Gamma^{\hat{j}}_{d}\geq\gamma_{th}\big]$  and $\Gamma_{d_{1}}\geq\gth$ implies that $\Gamma_{d}\geq\gth$, $q_1$ can be written as:
{\begin{eqnarray*}
q_1=\Pr\left[\Gamma^{\hat{j}}_{d}\geq\gamma_{th}, \Gamma_{r}\geq\gamma_{th}\right]-\underbrace{\Pr\left[\Gamma_{d_1}\geq\gamma_{th}, \Gamma_{r}\geq\gamma_{th}\right]}_{p_3}.
\end{eqnarray*}}
$q_1$ can be alternatively represented  in terms of $p_1$ and $p_2$ (and therefore  $p$ - see (\ref{eq:pout})) by using the $\Pr\big[\Gamma^{\hat{j}}_{d}\geq\gamma_{th}, \Gamma_{r}\geq\gamma_{th}\big]=1-\Pr\big[\Gamma_r < \gamma_{th}\big]-\Pr\big[\Gamma^{\hat{j}}_d< \gamma_{th},\;\Gamma_r \geq \gamma_{th}\big]$ as:
\begin{IEEEeqnarray}{rcl}
q_1&=&1-\bigg(\underbrace{\Pr\left[\Gamma_r< \gamma_{th}\right]}_{p_1}+\underbrace{\Pr\left[\Gamma^{\hat{j}}_d< \gamma_{th},\;\Gamma_r \geq \gamma_{th}\right]}_{p_2}+\underbrace{\Pr\left[\frac{P_s \hsd}{N_o}\geq\gamma_{th}, \frac{\xi P_s\bhsr}{N_o} \geq\gamma_{th} \right]}_{p_3}\bigg).\nonumber
\end{IEEEeqnarray}
From \eqref{eq:pout}, the above equation can be rewritten as:
$q_1=1-p-p_3$.
 We evaluate $p_3$ in the above expression, by integrating it first w.r.t $\bhsr
$ and $|h_{sd}|^{2}$ while conditioning it on $|g_{sp}|^{2}$ to get:
\begin{equation}
p_3\Big|_{\gsp}\hspace{-0.15cm}= \sum_{k=0}^{L-1}\frac{1}{k!}\left(\frac{\lsr\psi \gsp}{\xi}\right)^k e^{-\left(\frac{\lsr \psi}{\xi}+\lsd \psi\right) \gsp}\nonumber
\end{equation}
Now after averaging over $\gsp$, we get:
\begin{eqnarray}
p_3&=&\sum_{k=0}^{L-1}\frac{\lsp}{k!} \left(\frac{\lsr \psi}{\xi}\right)^k\int_{0}^{\infty}{(\gsp)}^k \exp\left(-\left(\frac{\lsr \psi}{\xi}+\lsd \psi +\lsp\right)\gsp\right)d\gsp\nonumber\\
&=&\frac{\xi\lsp}{\lsr \psi}\sum_{k=0}^{L-1} \left(\frac{{\lsr \psi}/{\xi}}{\left(\lsp+\lsd\psi+{\lsr\psi}/{\xi}\right)}\right)^{k+1}=\frac{1}{\frac{\lsd\psi}{\lsp}+1}\left(1-\left(\frac{\lsd \xi}{\lsr}+\frac{\lsp \xi}{\lsr \psi}+1\right)^{-L}\right).\label{eq:p3}
\end{eqnarray}
\begin{lemma}\label{lem:V4}
	The throughput $\tau_{in}$ of the incremental scheme is related to throughput of the cooperative cognitive scheme as follows:
	\begin{IEEEeqnarray}{rcl}
		\tau_{in}&=& \tau-0.5 \zeta R_s p_3+\frac{\zeta R_s}{1+\frac{\lsd \psi }{\lsp }}=\tau+\frac{0.5\zeta R_s}{1+\frac{\lsd \psi }{\lsp }}\bigg(1+\left(1+\frac{\lsd\xi}{\lsr}+\frac{\lsp\xi}{\lsr\psi}\right)^{-L}\bigg).\label{eq:tauIn_2}
	\end{IEEEeqnarray} 
\end{lemma}
\begin{IEEEproof}
	Using $q_1=1-p-p_3$ in (\ref{eq:tauIn}), and using (\ref{eq:tau_mrc}), we have:  $\tau_{in}=\zeta(\tau-0.5R_s p_3 + R_s q_2)$. Since $\Gamma_{d_1}=\frac{P_s\hsd}{N_o}=I \frac{\hsd}{\gsp}$, using the CDF of ratio of exponentials, it can be readily established that  $q_2=\Pr[\Gamma_{d_1}\geq\gamma_{th}]
	=\frac{1}{1+\frac{\lsd \psi }{\lsp }}$. Using this, along with the expression for $p_3$ in (\ref{eq:p3}), we have \eqref{eq:tauIn_2}.
\end{IEEEproof}
Denote by $\tau_{dir}$ the throughput of the link when the relay is absent (direct communication). Clearly, $\tau_{dir} = R_s\Pr\left[\frac{P_s\hsd}{N_o}\geq\gth\right]=\frac{R_s}{1+{\lsd \psi }/{\lsp }}$.
\begin{lemma}\label{tauin_tau}
	When the number of antennas at R is large, the asymptotic throughput difference ($\tau_{in}-\tau$) saturates to a fixed value equal to $0.5 \zeta\tau_{dir}$.
\end{lemma}
\begin{IEEEproof}
Using the fact that $\frac{\psi }{\lsp}\ll1$ and $\lsp\gg\lsd$ typically, which makes $\left(1+\frac{\lsd\xi}{\lsr}+\frac{\lsp\xi}{\lsr\psi}\right)^{-1}\ll1$, the above  can be approximated as:
	\begin{equation}
		\tau_{in}-\tau \hspace{0.1cm}\overset{L\rightarrow\infty}{=} \hspace{0.1cm}\frac{0.5 \zeta R_s}{1+\frac{\lsd \psi }{\lsp }}=0.5\zeta\,\tau_{dir}.\label{eq:DiffTauInAndTauMRCAsymptotic}
	\end{equation}
\end{IEEEproof}
On the other hand, for smaller number of antennas, the direct path contributes significantly to the throughput, and hence $\tau_{in}$ is much larger than $\tau$.
\subsubsection{Optimum EH parameter}
For both PS-EH and TS-EH, choice of optimal EH parameter $\rho$ for IR is of interest. We first show that $\tau_{in}$ is a concave function of $\rho$ in the following lemma.
\vspace{-0.3cm}
\begin{lemma}\label{Lem:TauINC_Concave}
	Throughput $\tau_{in}$ of the IR scheme is a concave function of $\rho$.
\end{lemma}
\vspace{-0.1cm}
\begin{IEEEproof}
	Proof is presented in Appendix-F.
\end{IEEEproof}
\begin{lemma}\label{Lem:rho_ts_in}
	For incremental relay case with $L=1$, optimum value $\rho_{in-ps}^{*}$ of $\rho$ equals $\rho_{ps}^{*}$. For TS-EH, optimum value of $\rho$, denoted by $	\rho^*_{in-ts}$,  is given by:
	\begin{subequations}\label{eq:rho_opt_in}
		\begin{IEEEeqnarray}{rcl}
		\rho^*_{in-ps}&\approx&  \rho^*_{ps}\label{eq:rho_ps_in}\\
		\rho^*_{in-ts}&\approx&\dfrac{1}{{2 \eta \lsp}/({\lrd \lsr \psi})-1}\bigg(-1+\dfrac{\eta }{\lrd \lsr}\sqrt{\dfrac{2 \lsd \lsp/\psi}{ {\eta \lsd}/({\lrd \lsr})+{2 \eta \lsp}/(\lrd \lsr \psi)-1}}\bigg)\,.\label{eq:rho_ts_in}
		\end{IEEEeqnarray}	
	\end{subequations}
\end{lemma}
\begin{IEEEproof}
Proof is presented in Appendix-G.
\end{IEEEproof}
\section{Simulation Results}\label{sec:simulations}
We validate the derived expressions using computer simulations. The normalized S-R, R-D,  S-P and R-P distances are chosen to be $1.2$, $1.8$, $3$ and $3$ units respectively. We assume  $\epsilon=4$ (path loss exponent), $\eta = 0.7$ and $I/N_o=6$ dB.
\par
\begin{figure*}[t]
	\vspace{-0.5cm}
	\begin{multicols}{2}
	\includegraphics[height=6.5cm,width=0.46\textwidth]{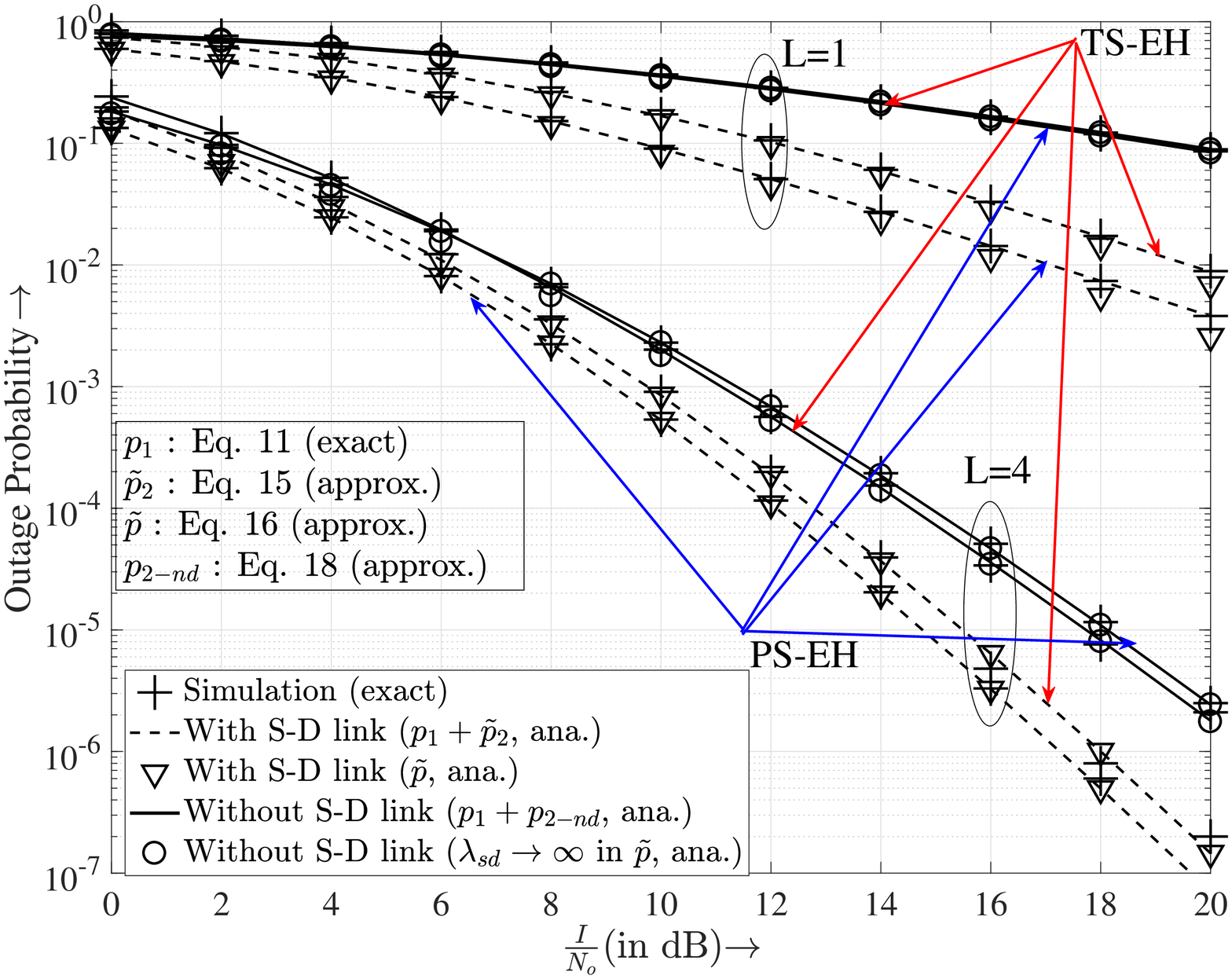}
	\par\caption{Outage probability vs. $\frac{I}{N_o}$ for different $L$}
	\label{fig:PoutVsIdB}
	\includegraphics[height=6.5cm,width=0.46\textwidth]{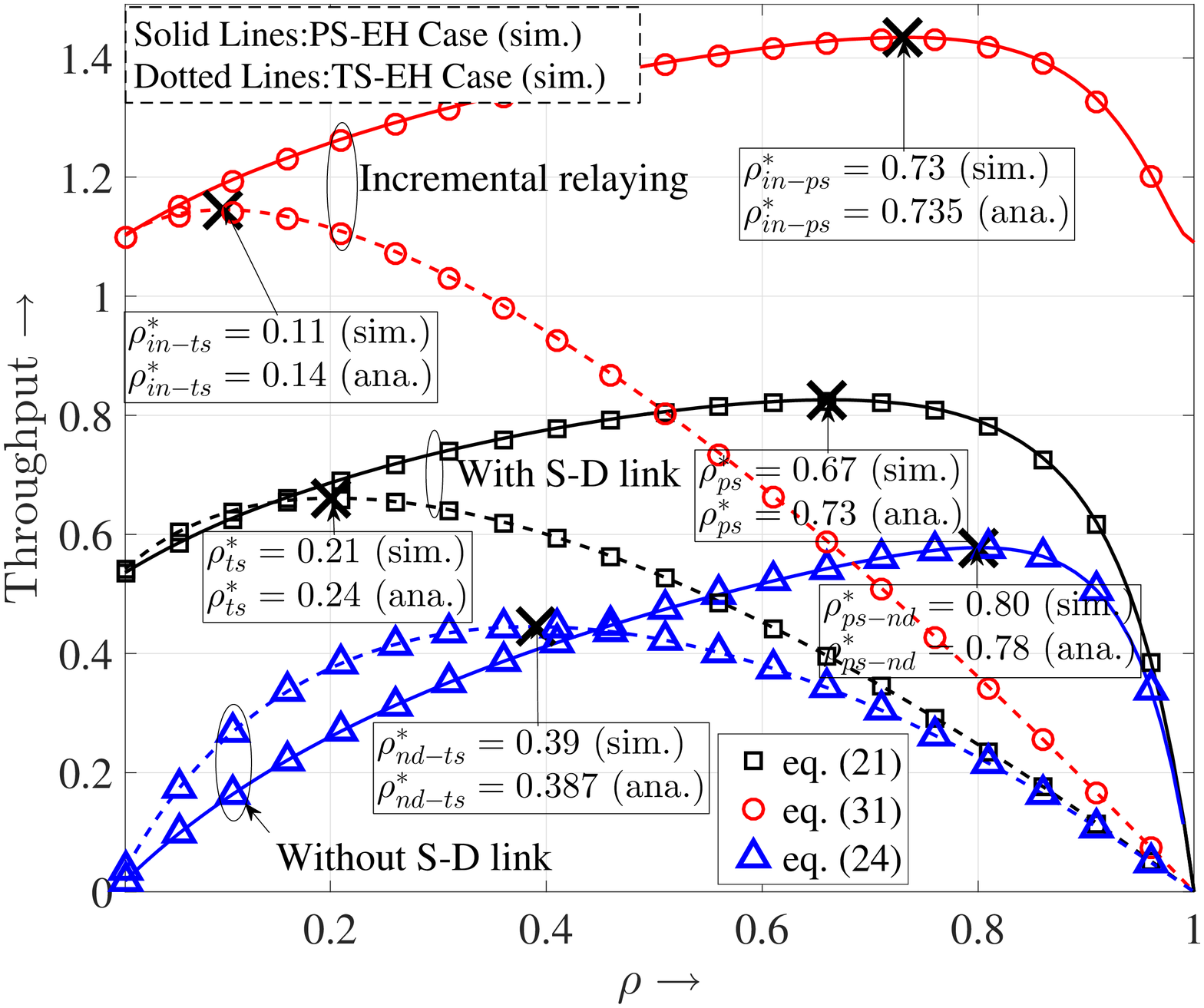}
	\par \caption{Throughput vs. $\rho$ for $L=1$, $R_s=3$ and $d_{sr}=1.5$}
	\label{fig:TauVsRhoForLEq1}	
	\end{multicols}
\vspace{-0.8cm}
\end{figure*}

 In Fig.~\ref{fig:PoutVsIdB}, outage performance of the EH-CCRN scheme (with and without the S-D channel) is depicted versus $I/N_o$ for $\rho=0.4$ and $R_s=1$ bits per channel use (bpcu). It can be observed that outage probability decreases with increase in number of antennas $L$. This is because the harvested energy as well as SNR at R increase significantly as  $L$ increases. Also, increase in degrees of freedom in choice of best channel from R to D boosts the second hop SNR further. Fig.~\ref{fig:PoutVsIdB} also verifies accuracy of the approximated outage probability expression in \eqref{eq:p2_WtRP}, and the approximation of \eqref{eq:pHSNR}.
\par Concavity of throughput with respect to $\rho$ (Lemma-\ref{lemma:TauConcaveWRTRho}), as well as the importance of choosing $\rho$ carefully are clearly brought out in Fig.~\ref{fig:TauVsRhoForLEq1} { with} $R_s=3$ bpcu. If $\rho$ is very small, the amount of harvested energy at R is not sufficient to support R-D transmission with both TS-EH and PS-EH protocols. On the other hand if it is too large ($\rho\rightarrow 1$), decoding of message signal at R becomes extremely difficult {with} PS-EH, and there is insufficient time for information transmission {with} TS-EH.  Similar arguments hold for throughput without the S-D path. It can be readily observed from the plot that the value of $\rho$ for IR protocol in the PS-EH case is almost the same as in EH-CCRN (Lemma ~\ref{Lem:rho_ts_in}),  and $\tau_{in}-\tau$ is approximately a constant (Lemma ~\ref{tauin_tau}). Optimum values obtained from the simulation are seen to closely matched the corresponding derived analytical optimum values. Also, the  optimum value of $\rho$ for TS-EH case is smaller than the optimum $\rho$ for the PS-EH case, and use of the S-D link results in smaller optimum $\rho$ values. Further, we verify the expression for optimum EH parameters derived in the paper for PS-EH and TS-EH { for} EH-CCRN (with and without S-D link from \eqref{eq:rho_opt} and \eqref{eq:rho_opt_withoutDSD} respectively) and IR from \eqref{eq:rho_opt_in}.
\par
\begin{figure*}[t]
	\begin{multicols}{2}
		\includegraphics[height=6.5cm,width=0.46\textwidth]{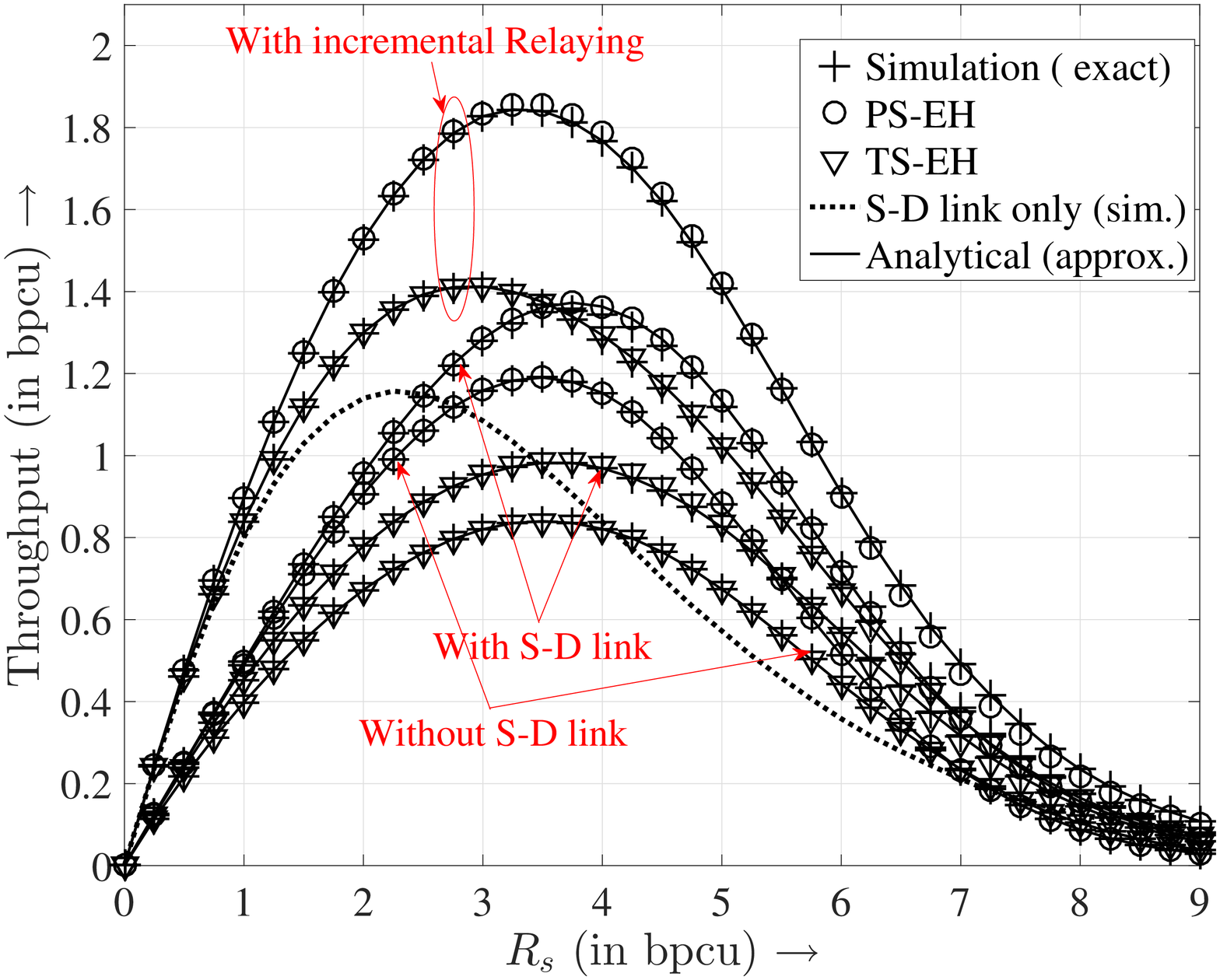}
		\par\caption{Throughput vs. $R_s$ for $L=2$ and optimum $\rho$}
		\label{fig:TauVsRsForLEq2}
		\includegraphics[height=6.5cm,width=0.46\textwidth]{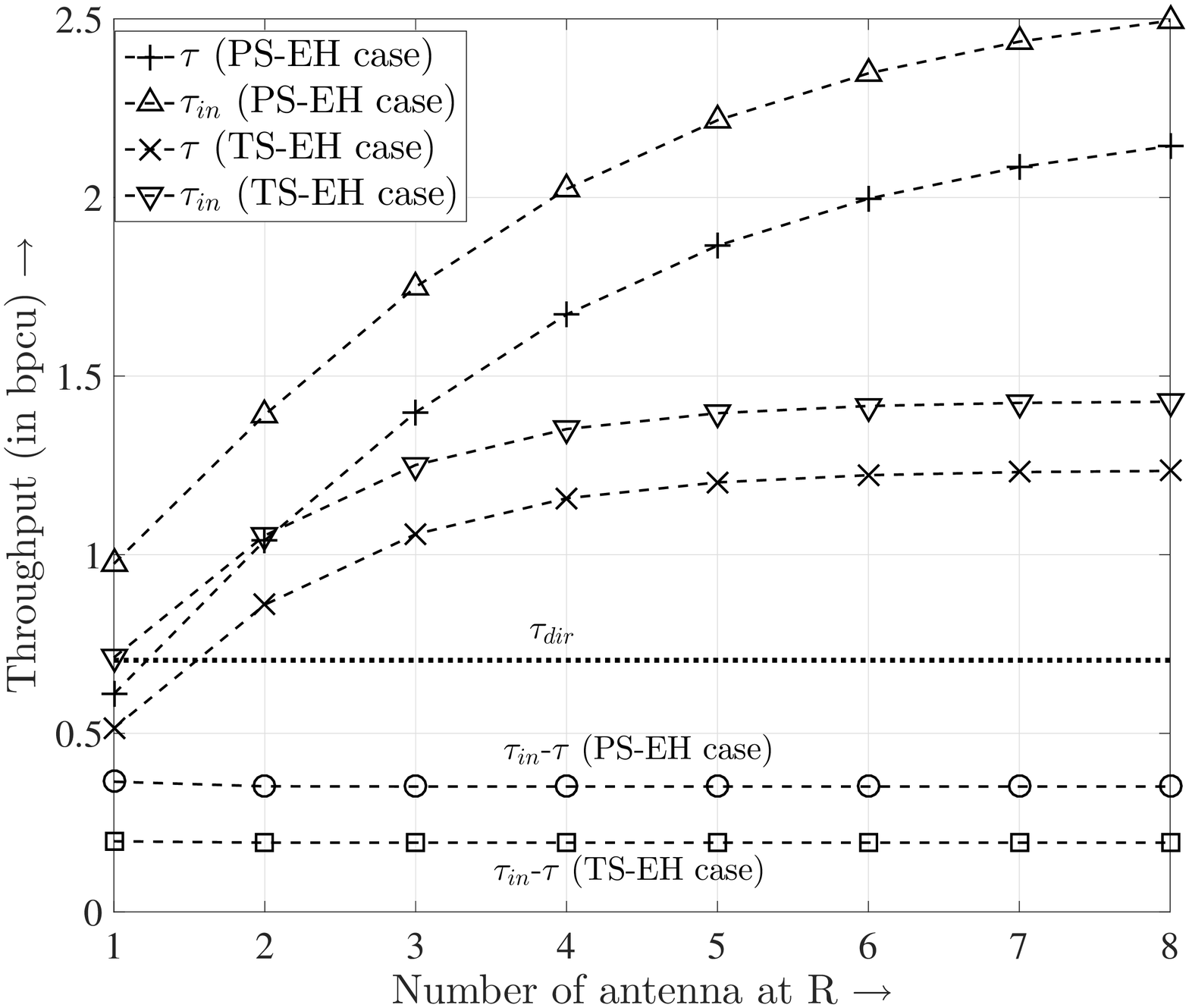}
		\par\caption{Throughput vs. different number of antenna at R}
		\label{fig:TauVsLForFixedRhoRs}	
	\end{multicols}
\vspace{-0.8cm}
\end{figure*}

 In Fig. \ref{fig:TauVsRsForLEq2}, SN throughput is plotted versus fixed  source rate ($R_s$) assuming $L=2$ {for optimum $\rho$ (which is determined by simulation).} For lower rate $R_s$, outage probability of the end-to-end communication is small and the number of bits received is also small. On the other hand, if $R_s$ is too high, end-to-end outage is also high and this again lowers the number of bits received by D. Hence, $R_s$ needs to be optimally chosen for maximum throughput in all the cases (TS-PH or PS-EH, EH-CCRN and IR). 
\par In Fig. \ref{fig:TauVsLForFixedRhoRs}, we plot the throughput versus number of antennas $L$ at R for $\rho=0.4$ and $R_s=4.5$. It is evident that increase in $L$ improves throughput. Also, IR always results in better throughput even when $L=1$ (Lemma \ref{lem:V4}). 
The difference between $\tau_{in}-\tau$ is a constant quantity as evident from \eqref{eq:DiffTauInAndTauMRCAsymptotic}. { It is also evident that performance is inferior to the case when only the direct (relay-less link) is used when $L=1$. This clearly demonstrates that use of more than one antenna is important.}
\begin{figure}[h]
	\vspace{-0.3cm}
        \centering
        \includegraphics[height=6.5cm,width=0.5\textwidth]{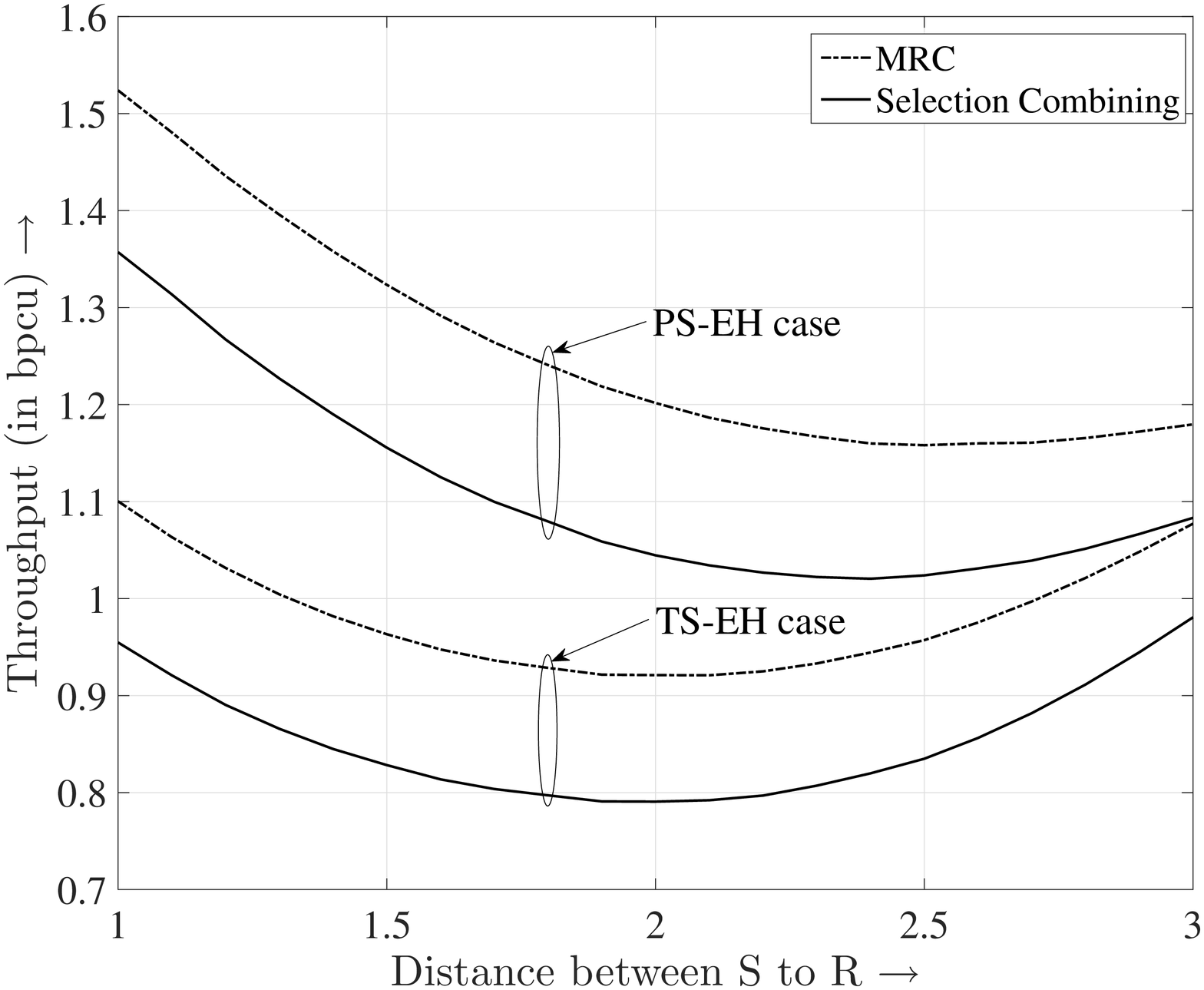}
	    \caption{Far-user throughput (in bpcu) vs. normalized distance between S to R for $L=2$, and TS-EH and PS-EH cases }
	    \label{fig:TauVsdSR_Leq2}
	    \vspace{-0.3cm}	 
\end{figure}
\par  In Fig. \ref{fig:TauVsdSR_Leq2}, throughput is plotted vs. distance between nodes S and R for the following parameters: $L=2, R_s = 4, I/N_o= 9$ dB. Here the normalized S-D, S-P, and R-P distances are 4. R-D distance is equal to the difference between S-D and S-R distances. Recently \cite{Xie2018} investigated the performance of EH-relayed underlay-CRN with the selection combining (SC). Clearly, unlike the MRC scheme considered in this paper, is not an optimal transmission scheme. We can observe a throughput gain of more than 0.1 bpcu when  MRC is used. 
\vspace{-0.3cm}
\section{Conclusion} \label{sec:conclusion}
In this paper, we analyzed throughput performance of a cooperative cognitive two-hop network with a multi-antenna energy harvesting relay for both time-switching and power-splitting protocols. Unlike other works to date, we {optimally combine the direct and relayed signals}. We demonstrated that the direct channel contributes significantly to throughput when the number of antennas at the energy harvesting relay is small. Incremental relaying was shown to result in large throughput gains. We presented closed form expressions for optimum values of charging parameters in several important cases, and derived insights into performance of such systems.

\section*{Appendix-A:Derivation of approximation of $p_2$}
\setcounter{equation}{0}
\renewcommand{\theequation}{A.\arabic{equation}}
From \eqref{eq:p2_def}, $p_2$ can be rewritten as follows:
{\begin{equation}
	p_2=\Pr\bigg[\frac{P_s \hsd}{N_o}+\frac{X\big|_{P_s\bhsr}}{N_o}< \gamma_{th},\frac{\xi P_s \bhsr}{N_o} \geq \gamma_{th}\bigg].\label{eq:p2Appendix}
\end{equation}}
where $X={\displaystyle\max_{j}}\left(\min\left(^hP_r, {I}/{|g^j_{rp}|^2}\right)|h^j_{rd}|^2\right)$.  To find a CDF of X we first write the CDF of $X_j = \min\left(^hP_r, {I}/{|g^j_{rp}|^2}\right)|h^j_{rd}|^2$, and then use order statistics to write CDF of $X$ conditioned on $^hP_r$ as \cite{Papoulis2002}:
{	\begin{IEEEeqnarray}{rcl}
		F_{X\big|_{^hP_r}}(x) &=& \left(F_{X_j}\left(x_j\Big|\,^hP_r\right)\right)^L\,, \label{eq:CDF_X1}
\end{IEEEeqnarray}}	
where $F_{X_j}\left(x_j\Big|\,^hP_r\right)$ is the CDF of $X_j$ conditioned on $\,^hP_r$, which can be evaluated as follows:
\begin{IEEEeqnarray*}{rcl}
	F_{X_j}\left(x_j\Big|\,^hP_r\right)&=& \Pr[X_j<x_j\Big|\,^hP_r]\nonumber\\
	&=& \Pr\left[^hP_r|h^j_{rd}|^2< x_j,\, ^hP_r\leq {I}/{|g^j_{rp}|^2}\right]+ \Pr\left[{I|h^j_{rd}|^2}/{|g^j_{rp}|^2}< x_j,\, ^hP_r> {I}/{|g^j_{rp}|^2}\right]\,.
\end{IEEEeqnarray*}
By utilizing the independence between $|h^j_{rd}|^2$ and $|g^j_{rp}|^2$, we have:
{\begin{IEEEeqnarray}{rcl}	
		F_{X_j}\left(x_j\Big|\,^hP_r\right)=1-e^{\frac{\lambda_{rd}x}{\beta\,P_s\,\bhsr}}\Big(1-\frac{\frac{\lambda_{rd} x}{I\lambda_{rp}} e^{-\frac{\lambda_{rp}I}{\beta\,P_s\,\bhsr}}}{1+\frac{\lambda_{rd}x}{I\lambda_{rp}}}\Big).\nonumber
\end{IEEEeqnarray}}		
where $^hP_r=\beta P_s \parallel {\bf h}_{sr} \parallel^{2}$. From \eqref{eq:CDF_X1} using the binomial theorem, CDF of $X$ conditioned on $^hP_r$ can be expressed as:
\vspace{-0.6cm}
\begin{IEEEeqnarray}{rcl}			
	F_{X}(x\big|{^hP_r})&=&\sum_{i=0}^{L} \sum_{j=0}^i\binom{L}{i}\binom{i}{j}(-1)^{i+j}\left(1-\frac{1}{1+\frac{\lambda_{rd}x}{I\lambda_{rp}}}\right)^{j} \exp\left(-\frac{j\lambda_{rp}I}{\beta\,P_s\,\bhsr}-\frac{i\lambda_{rd}x}{\beta\,P_s\,\bhsr}\right). \label{eq:CDF_X}
\end{IEEEeqnarray}
After averaging \eqref{eq:p2Appendix} over $X$, $p_2$ becomes:
\begin{IEEEeqnarray}{rcl}
	p_2&=&\E_{\C_1  }\left[	F_{X\big|{^hP_r}}\left(\gamma_{th}-\frac{P_s \hsd}{N_o}\right)\right],\nonumber
\end{IEEEeqnarray}
where the condition $\C_1=\big\{{\xi P_s \bhsr}\geq\psi,\, \frac{P_s \hsd}{N_o}<\gth\big\}$. Using the CDF of $X$ from \eqref{eq:CDF_X}, the above equation is expanded as follows: 
\begin{IEEEeqnarray}{rcl}
	p_2&=&\sum_{i=0}^{L} \sum_{j=0}^i\binom{L}{i}\binom{i}{j}(-1)^{i+j}\E_{\C_1  }\Big[e^{-\frac{i\lambda_{rd}\left(\gamma_{th}N_o-P_s \hsd\right)}{\beta\,P_s\,\bhsr}}e^{-\frac{j\lambda_{rp}I}{\beta\,P_s\,\bhsr}}\Big(1-\frac{1}{1+\frac{\lambda_{rd}}{I\lambda_{rp}}(\gamma_{th}N_o-P_s \hsd)}\Big)^{j}\Big],\nonumber
\end{IEEEeqnarray}
 Here,  $\frac{\lambda_{rd}}{I\lambda_{rp}}(\gamma_{th}N_o-P_s \hsd)$ that appears in the denominator is such that it is much less than one. For this reason, to make the analysis tractable, $\frac{\lambda_{rd}}{I\lambda_{rp}}\left(\gamma_{th}N_o-P_s \hsd\right)$ {is} replaced by its mean conditioned on $\C_1$  i.e. $\frac{\lambda_{rd}}{I\lambda_{rp}}\left(\gamma_{th}N_o-\E_{\C_1}[P_s \hsd]\right)$. $\E_{\C_1}[P_s \hsd]$ can be derived as:
\begin{IEEEeqnarray}{rcl}
\E_{\C_1}[P_s \hsd]&=&\frac{I \lambda_{sp}}{\lambda_{sd}} \Big(\log \Big(\frac{\psi \lambda_{sd}}{\lambda_{sp}}+1\Big)+\frac{\psi \lambda_{sd}}{\lambda_{sp}}\frac{1}{1+\frac{\psi \lambda_{sd} }{\lambda_{sp}}}\Big).\nonumber
\end{IEEEeqnarray}
With this, $p_2$ can be represented in approximated form as: 
\begin{IEEEeqnarray}{rcl}
	p_2&\approx&\sum_{i=0}^{L} \sum_{j=0}^i\binom{L}{i}\binom{i}{j}(-1)^{i+j}t^j\E_{\C_1}\Big[{e^{-\frac{i\lambda_{rd}\left(\gamma_{th}N_o-P_s \hsd\right)}{\beta\,P_s\,\bhsr}}}e^{-\frac{j\lambda_{rp}I}{\beta\,P_s\,\bhsr}}\Big],
\end{IEEEeqnarray}
where $t=1-1/({1+\frac{\lambda_{rd}}{I\lambda_{rp}}\left(\gamma_{th} N_o-\E_{\C_1}\left[P_s\hsd\right]\right)})$. The above approximation is tight for $\lrd\ll \lrp$. This is true for general system settings in underlay-CRN. Now averaging the above equation w.r.t. $\hsd$, $p_2$ becomes:
\begin{IEEEeqnarray}{rcl}
	p_2&\approx&\sum_{i=0}^{L} \sum_{j=0}^i\binom{L}{i}\binom{i}{j}(-1)^{i+j}t^j\E_{\C_3}\Big[e^{-\frac{j\lambda_{rp}I}{\beta\,P_s\,\bhsr}}\int_{0}^{\frac{\psi}{P_s}}\hspace{-0.1cm}{e^{-\frac{i\lambda_{rd}\left(\gamma_{th}-P_s \hsd\right)}{\beta\,P_s\,\bhsr}}}\lsd e^{-\lsd \hsd}\Big],\nonumber
\end{IEEEeqnarray}
where condition $\C_3=\left\{{\xi P_s \bhsr}/{N_o}>\gth\right\}$. Performing the integrations, we can write $p_2$ as:
\begin{equation}
	p_2=\sum_{i=0}^{L} \sum_{j=0}^i\binom{L}{i}\binom{i}{j}(-1)^{i+j}t^j\E_{\C_3}\bigg[\frac{\beta \bhsr \lambda_{sd} }{\beta \bhsr \lambda_{sd}-i \lambda_{rd}} e^{-\frac{I j \lambda_{rp}}{\beta \bhsr P_s}} \Big(e^{-\frac{i \lambda_{rd}\gamma_{th} N_o}{\beta \bhsr P_s}}-e^{-\frac{\lambda_{sd}\gamma_{th} N_o}{P_s}}\Big)\bigg],\label{eq:p_2_hsrgsp}
\end{equation}
Substituting the value of $P_s=I/\gsp$ from \eqref{eq:PsOnlyInterference}, \eqref{eq:p_2_hsrgsp} becomes:
{\begin{equation}
		p_2\approx\sum_{i=0}^{L} \sum_{j=0}^i\binom{L}{i}\binom{i}{j}(-1)^{i+j}t^j\E_{\C_3}\Big[\frac{\beta \bhsr \lambda_{sd}}{\beta \bhsr \lambda_{sd}-i \lambda_{rd}} e^{-\frac{j \lambda_{rp}\gsp}{\beta \bhsr}} \Big(e^{-\frac{i \lambda_{rd}\psi\gsp}{\beta \bhsr }}-e^{-{\lambda_{sd}\psi\gsp}}\Big)\Big].\label{eq:p_2_hsrgsp1}
\end{equation}}
To make the expression compact, we use: $a=\frac{\lambda_{rp}}{\beta}$, $b=\frac{\psi  \lambda_{rd} }{\beta}$, $c={\psi\lambda_{sd}}$, $d=\frac{\lambda_{rd}}{\beta \lambda_{sd}}$, and $g=\frac{\xi}{\psi}$. After averaging over $\gsp$, the above equation becomes:
{\begin{IEEEeqnarray}{rcl}
		&&p_2\approx\sum_{i=0}^{L} \sum_{j=0}^i\binom{L}{i}\binom{i}{j}(-1)^{i+j}t^j\E_{\bhsr}\Big[\frac{\bhsr}{\bhsr-i d}\nonumber\\
		&&\mathlarger{\int}_0^{g \bhsr}\hspace{-0.7cm} \lsp e^{-\frac{j a \gsp}{\bhsr}} e^{ -\lambda_{sp}\gsp} \Big(e^{-\frac{i b \gsp}{\bhsr}}-e^{-c \gsp}\Big) \,  d\gsp\Big]. \nonumber
\end{IEEEeqnarray}}
After integration, $p_2$ conditioned on $\bhsr$ can now be expressed as:
{{\small\begin{IEEEeqnarray}{rcl}
		&&p_2\big|_{\bhsr}=\sum_{i=0}^{L} \sum_{j=0}^i\binom{L}{i}\binom{i}{j}(-1)^{i+j}t^j \frac{(\bhsr)^2 \lambda_{sp}}{\bhsr-i d}\Bigg(\frac{1-e^{-g (j a+i b+\bhsr \lambda_{sp})}}{j a+i b+\bhsr \lambda_{sp}}+\frac{e^{-g (j a+\bhsr (c+\lambda_{sp}))}-1}{j a+\bhsr (c+\lambda_{sp})}\Bigg).\nonumber
	\end{IEEEeqnarray}}
After some manipulations, the above equation can be represented as in {\eqref{eq:p2_2} using $b=cd$.}
	{\begin{IEEEeqnarray}{rcl}
			p_2|_{\bhsr}&=&\sum_{i=0}^{L} \sum_{j=0}^i\binom{L}{i}\binom{i}{j}(-1)^{i+j}t^j\Bigg\{\frac{1}{j a+i b+i d \lsp}\Bigg(\frac{j a \lambda_{sp} \left(1-e^{-g (j a+\bhsr (c+\lambda_{sp}))}\right)}{(c+\lambda_{sp})(1+ \frac{(c+\lambda_{sp})}{j\,a}\bhsr)}\nonumber\\
			&&+i \,d \lambda_{sp}e^{-g j a}e^{-g \lambda_{sp}\bhsr}\bigg(\frac{e^{-g\,c\bhsr}}{\frac{\bhsr}{i d}-1}-\frac{e^{-g i b}}{\frac{\bhsr}{id}-1}\bigg)-\frac{(j a+i b) \left(1-e^{-g (j a+i b+\bhsr \lambda_{sp})}\right)}{(1+\frac{\lsp}{j a+i b}\bhsr)}\Bigg)\nonumber\\
			&&-e^{-g (j a+i b+\bhsr \lambda_{sp})}+\frac{\lambda_{sp} e^{-g (j a+\bhsr (c+\lambda_{sp}))}}{c+\lambda_{sp}}+\frac{c}{c+\lambda_{sp}}\Bigg\}.	\label{eq:p2_2}
	\end{IEEEeqnarray}}
{ Finally, averaging \eqref{eq:p2_2} over $\bhsr$, $p_2$ is given by: 
{\small	\begin{IEEEeqnarray}{rcl}
		p_2&=&\sum _{i=0}^L \binom{L}{i} \sum _{j=0}^i (-1)^{i+j} t_1^j \binom{i}{j} \int_0^{\infty } \frac{(\bhsr)^{L-1} \lsr^L}{(L-1)!}e^{-\lsr\bhsr}  \Bigg(\frac{a^2 \lsp j^2 \left(1-e^{-g (a j+\bhsr (c+\lsp))}\right)}{(c+\lsp) (a j+\bhsr (c+\lsp)) (a j+b i+d \lsp i)}\nonumber\\
		&&\frac{\lsp (d i)^2 }{a j+b i+d \lsp i}\left(\frac{e^{-g (a j+\bhsr (c+\lsp))}}{\bhsr-d i}-\frac{e^{-g (a j+b i+\bhsr \lsp)}}{\bhsr-d i}\right)-\frac{(a j+b i)^2 \left(1-e^{-g (a j+b i+\bhsr \lsp)}\right)}{(a j+b i+d \lsp i) (a j+b i+\bhsr \lsp)}\nonumber\\
		&&-e^{-g (a j+b i+\bhsr \lsp)}+\frac{\lsp e^{-g (a j+\bhsr (c+\lsp))}}{c+\lsp}+\frac{c}{c+\lsp}\Bigg) \, d\bhsr .\label{eq:p2_3}
	\end{IEEEeqnarray}}
Using the relation $\frac{x^L}{x - a} = 
\sum_{k=0}^{L-1}a^k x^{L - 1 - k} + \frac{a^L}{x - a}$, $p_2$ can be rewritten as follows:
{\small	\begin{IEEEeqnarray}{rcl}		
		p_2&=&\sum _{i=0}^L \binom{L}{i} \sum _{j=0}^i (-1)^{i+j} t_1^j \binom{i}{j} \Bigg(\frac{\lsr^L \left(\lsp (d i)^2\right) }{(L-1)! (a j+b i+d \lsp i)}\bigg(\int_0^{\infty } \frac{\lsr^L (d i)^{L-1} e^{-\lsr\bhsr}}{(L-1)! (\bhsr-d i)} \big(e^{-g (a j+\bhsr (c+\lsp))}\nonumber\\
		&&-e^{-g (a j+b i+\bhsr \lsp)}\big) \, d\bhsr+\sum _{k=0}^{L-2} (d i)^k \Gamma (-k+L-1) \Big(e^{-a g j} (g (c+\lsp)+\lsr)^{k-L+1}-e^{-g (a j+b i)}\nonumber\\
		&& \times(g \lsp+\lsr)^{k-L+1}\Big)\bigg)+\frac{(a j)^2}{(c+\lsp)^2 (a j+b i+d \lsp i)}\int_0^{\infty } \frac{\lsp e^{-\lsr\bhsr} (\bhsr)^{L-1} \lsr^L}{(L-1)! \left(\frac{a j}{c+\lsp}+\bhsr\right)} \nonumber\\
		&&\times\left( \left(1-e^{-g (a j+\bhsr (c+\lsp))}\right)\right) \, d\bhsr-\frac{(a j+b i)^2\lsr^L}{(a j+b i+d \lsp i)(\lsp (L-1)!)}\int_0^{\infty } \frac{e^{-\lsr\bhsr} (\bhsr)^{L-1}}{ \left(\frac{a j+b i}{\lsp}+\bhsr\right)} \nonumber\\
		&&\left(1-e^{-g (a j+b i+\bhsr \lsp)}\right)\ d\bhsr-\int_0^{\infty } \frac{e^{-\lsr\bhsr}(\bhsr)^{L-1} \lsr^Le^{-g (a j+b i+\bhsr \lsp)}}{(L-1)!} \, d\bhsr\nonumber\\
		&&+\int_0^{\infty } \frac{e^{-\lsr\bhsr} (\bhsr)^{L-1} \lsr^L \left(\lsp e^{-g (a j+\bhsr (c+\lsp))}\right)}{(L-1)! (c+\lsp)} \, d\bhsr+\frac{c}{c+\lsp}\Bigg).\label{eq:p2_5}
	\end{IEEEeqnarray}}
	To solve the integrals in \eqref{eq:p2_5}, we use the following relations:
	\begin{IEEEeqnarray}{rcl}
		\int_0^{\infty } \frac{x^{L-1} \exp (-b x)}{a+x} \, dx&=&e^{a b} b^{1-L} \Gamma (L) E_L(a b)\qquad\text{and}\nonumber\\
		\int_0^{\infty } e^{-\lsr\bhsr} \left(\frac{e^{-g (\bhsr (c+\lsp))}}{\bhsr-d i}-\frac{e^{-g (b i+\bhsr \lsp)}}{\bhsr-d i}\right) d\bhsr
		&=&e^{-i(b g+d (g \lsp+\lsr))} (\text{Ei}(i d (g \lsp+\lsr) )\nonumber\\
		&&\hspace{1cm}-\text{Ei}(i d (c g+\lsp g+\lsr))).\nonumber
	\end{IEEEeqnarray}
	The $p_2$ can finally be rewritten as follows:
{\small		\begin{IEEEeqnarray}{rcl}
			p_2&=&\sum _{i=0}^L  \sum _{j=0}^i (-1)^{i+j} t_1^j \binom{L}{i} \binom{i}{j} \bigg(\frac{\lsr^L \lsp (d i)^2}{(L-1)! (a j+b i+d \lsp i)} \Big(\sum _{k=0}^{L-2} (d i)^k \Gamma (-k+L-1) \Big(e^{-a g j} (g (c+\lsp)+\lsr)^{k-L+1}\nonumber\\
			&&-e^{-g (a j+b i)} (g \lsp+\lsr)^{k-L+1}\Big)+(d i)^{L-1} e^{-g (a j+b i)} e^{-d i (g \lsp+\lsr)} (\text{Ei}(d (g \lsp+\lsr) i)-\text{Ei}(d (c g+\lsp g+\lsr) i))\Big)\nonumber\\
			&&+\frac{\lsp \lsr e^{a_2 \lsr}}{a j+b i+d \lsp i} \left(\frac{a j}{c+\lsp}\right)^2\left(E_L(a_2 \lsr)-\left(\frac{\lsr}{g (c+\lsp)+\lsr}\right)^{L-1} E_L(a_1 (g (c+\lsp)+\lsr))\right)\nonumber\\
			&&-\frac{\lsr}{\lsp (a j+b i+d \lsp i)} e^{a_1 \lsr} (a j+b i)^2 \left(E_L(a_1 \lsr)-\left(\frac{\lsr}{g \lsp+\lsr}\right)^{L-1} E_L(a_1 (g \lsp+\lsr))\right)-\left(\frac{\lsr}{g \lsp+\lsr}\right)^L \nonumber\\
			&&\times e^{-g (a j+b i)}+\frac{\lsp}{c+\lsp} e^{-a g j}\left(\frac{\lsr}{g (c+\lsp)+\lsr}\right)^L+\frac{c}{c+\lsp}\bigg)\,,\label{eq:p2_4}
		\end{IEEEeqnarray}}
where $b_1=e^{-g (a j+b i)}$, $a_1=\frac{a j+b i}{\lsp}$,  $b_2=e^{-a \,g\,j}$ and $a_2=\frac{a j}{c+\lsp}$.}
\section*{Appendix-B: {Approximation of $\tilde{p}$}}
\setcounter{equation}{0}
\renewcommand{\theequation}{B.\arabic{equation}}
\subsubsection*{Approximation of $\tilde{p}_2$}
{For the typical value of parameters ( i.e. $\lsp \gg \psi$)}, arguments inside $e^{-x}\text{Ei}(x)$ and $e^{x} \E_k(x), k>0$ increase and decrease respectively with increase in $x$. From the fact that $e^{-x}\text{Ei}(x)\rightarrow 0$ for $x\rightarrow\infty$, the term associated with $\text{Ei}(x)$ vanishes from the expression, and it can be expressed as follows:
{\small\begin{IEEEeqnarray}{rcl}
	\tilde{p}_2&\approx&\sum _{i=1}^L (-1)^i \binom{L}{i} \Bigg(\frac{\lrd^2 \lsp \lsr^L }{\beta \lsd^2 (L-1)! \left(\frac{\lrd \lsp}{ \lsd}+{\lrd \psi}\right)}\bigg(\sum _{k=0}^{L-2} i\, \Gamma (L-k-1) \left(\frac{\lrd i}{\beta \lsd}\right)^k \bigg(\left(\frac{\xi (\lsd \psi+\lsp)}{\psi}+\lsr\right)^{k-L+1}\nonumber\\
	&&-e^{-\frac{\lrd i \xi}{\beta}}\left(\frac{\lsp \xi}{\psi}+\lsr\right)^{k-L+1}\bigg)\bigg) -\frac{\left(\lrd^2 \lsr i \psi^2 \right)}{\beta \lsp \left(\frac{\lrd \lsp}{ \lsd}+{\lrd \psi}\right)}e^{\frac{\lrd \lsr i \psi}{\beta \lsp}}\bigg\{E_L\left(\frac{\lrd \lsr i \psi}{\beta \lsp}\right)-\left({\frac{\lsp \xi}{\psi \lsr}+\lsr}\right)^{1-L} \nonumber\\
	&& \times E_L\left(\frac{\lrd i \psi}{\beta \lsp} \left(\lsr+\frac{\lsp \xi}{\psi}\right)\right)\bigg\}-e^{-\frac{\lrd i \xi}{\beta}} \left({\frac{\lsp \xi}{\psi\lsr}+1}\right)^{-L}\Bigg)-\left({\frac{\lsp \xi}{\lsr \psi}+1}\right)^{-L}.
\label{eq:p2_WtRPWithoutEix}
\end{IEEEeqnarray}}
To further simplify \eqref{eq:p2_WtRPWithoutEix}, we use the fact that  $e^x E_L(x)>e^x E_L(x+A)$ for $A>0$. If $A$ is very large (as it is in this case since $A=\xi \lsp/\psi\gg0$ for $\rho>0$), the term with $e^x Ei(-(x+A))$ can be neglected. \eqref{eq:p2_WtRPWithoutEix} now becomes as follows:
	{\begin{IEEEeqnarray}{rcl}
		\tilde{p}_2&\approx&\sum _{i=1}^L (-1)^{i+1} \binom{L}{i} \Bigg(\underbrace{\frac{\lrd \lsr^L }{\beta \lsd (L-1)! \left(1+{\psi\lsd}/\lsp\right) \left(\frac{\lsp \xi}{\psi}+\lsr\right)^{L-1}}\bigg(\sum _{k=0}^{L-2} i\, \Gamma (L-k-1) \left(\frac{\lrd i}{\beta \lsd}\right)^k}_{T_1\ldots}\nonumber\\
		&&\underbrace{\times \bigg(e^{-\frac{\lrd i \xi}{\beta}}\left(\frac{\lsp \xi}{\psi}+\lsr\right)^{k}-\left(1+\frac{\xi \lsd}{\frac{\lsp \xi}{\psi}+\lsr}\right)^{1-L}\left(\frac{\xi (\lsd \psi+\lsp)}{\psi}+\lsr\right)^{k}\bigg)\bigg)}_{\ldots T_1}\nonumber\\
		&& +\frac{1}{\frac{ \lsp}{  \psi \lsd}+1}\frac{\lrd \lsr i \psi}{\beta \lsp}e^{\frac{\lrd \lsr i \psi}{\beta \lsp}}E_L\left(\frac{\lrd \lsr i \psi}{\beta \lsp}\right)+e^{-\frac{\lrd i \xi}{\beta}} \left(\frac{1}{\frac{\lsp \xi}{\psi\lsr}+1}\right)^L\Bigg)-\left(\frac{1}{\frac{\lsp \xi}{\lsr \psi}+1}\right)^L.\label{eq:tilde_p2_1}
	\end{IEEEeqnarray}}
{For $L=1$, the value of the term $T_1$ vanishes (as this term does not exist for $L=1$). For $L\geq2$, it can be shown that $T_1$ is very small compared to other terms as it is depend on $({\lsr+\xi\lsp/\psi})^{1-L}/(L-1)!$. After neglecting $T_1$ in the above equation, $\tilde{p}_2$ can further approximated as:}
\begin{equation}
		\tilde{p}_2\approx\sum _{i=1}^L (-1)^{i+1} \binom{L}{i} \Bigg(\frac{\frac{\lrd \lsr i \psi}{\beta \lsp}}{\frac{ \lsp}{  \psi \lsd}+1}e^{\frac{\lrd \lsr i \psi}{\beta \lsp}}E_L\left(\frac{\lrd \lsr i \psi}{\beta \lsp}\right)+e^{-\frac{\lrd i \xi}{\beta}} \left({\frac{\lsp \xi}{\psi\lsr}+1}\right)^{-L}\Bigg)
		-\left({\frac{\lsp \xi}{\lsr \psi}+1}\right)^{-L}.\label{eq:p2_WtRP_APP2}
\end{equation}
We note that above approximation may not {hold} for {very high} operating rate as assumption {$\psi\ll\lsp$} violates in this case. However, high source rate transmission increase the outage probability between S to D which ultimately lower the throughput. Due to this reason we generally operate with moderate source rate (for which { the} approximation work {well}) which maximizes the throughput.
\section*{Appendix-C:Proof of concavity of function $F_1\left(\rho_{ps}\right)$ and $F_2\left(\rho_{ps}\right)$ for PS-EH case}
\setcounter{equation}{0}
\renewcommand{\theequation}{C.\arabic{equation}}
\subsubsection*{Proof of concavity of function $F_1\left(\rho_{ps}\right)$}
From \eqref{eq:tauMRCLRPtendInfty}, and for PS-EH case $\beta=\eta\rho_{ps}$. Function $F_1\left(\rho_{ps}\right)$ is expressed as:
\begin{IEEEeqnarray}{rcl}
F_1(\rho_{ps})=\sum _{i=1}^L (-1)^{i+1}\binom{L}{i} L e^{\frac{\psi i \lrd \lsr }{\rho_{ps} \eta \lsp}} E_{L+1}\Big(\frac{\psi i \lrd \lsr }{\rho_{ps} \eta \lsp}\Big)-1.\nonumber
\end{IEEEeqnarray}
To prove concavity of the expression we have to show that $\frac{d^2 F_1(\rho_{ps})}{d\rho_{ps}^2}\leq0$ \cite{Boyd2004}. {For convenience}, we use the approximation of $F_1(\rho_{ps})$ (say $F'_1(\rho_{ps})$) to prove the concavity. For general system parameters i.e. ${\psi  \lrd \lsr  }/({\eta  \lsp})\ll1$, {the following approximation} hold  (from \cite[5.1.19]{Abramowitz1964}}.  
\renewcommand{\theequation}{C.\arabic{equation}}
\begin{eqnarray}
e^{\frac{\psi i \lrd \lsr }{\rho_{ps} \eta \lsp}} E_{L+1}\Big(\frac{\psi i \lrd \lsr }{\rho_{ps} \eta \lsp}\Big)\approx \frac{1}{L+\frac{\psi i \lrd \lsr }{\rho_{ps} \eta \lsp}}.\nonumber
\end{eqnarray}
Approximation of $F_1(\rho_{ps})$ is {then} given by:
\begin{IEEEeqnarray}{rcl}
F'_1(\rho_{ps})&=&\sum _{i=1}^L (-1)^{i+1}\binom{L}{i} \frac{L}{L+\frac{\psi i \lrd \lsr }{\rho_{ps} \eta \lsp}}-1 \,.\nonumber
\end{IEEEeqnarray}
From the relation (\cite[eq. 4.2.2.45]{Prudnikov1992}), {we have: }
\begin{eqnarray}
\sum_{i=1}^{L}(-1)^{i+1}\binom{L}{i}\frac{ix}{i x+1}= \frac{\Gamma(L+1)\Gamma(1+1/x)}{\Gamma(L+1+1/x)}.\label{eq:SumRel}
\end{eqnarray}
$F'_1(\rho_{ps})$ can be expressed as:
\begin{equation}
F'_1(\rho_{ps})=-\frac{\Gamma (L+1) \Gamma \left(\frac{\eta\rho_{ps} L \lsp}{\lrd \lsr \psi}+1\right)}{\Gamma \left(\frac{\eta\rho_{ps} L \lsp}{\lrd \lsr \psi}+L+1\right)}.\label{eq:FDash1}
\end{equation}
From the approximation of ratio of Gamma function 
\cite[1]{Tricomi1951} i.e.:
\begin{eqnarray}
\vspace{-0.4cm}
\frac{\Gamma[z+r]}{\Gamma[z+s]} \leq z^{r-s}\,. \label{eq:RatoOfGamma}
\vspace{-0.4cm}
\end{eqnarray}
\eqref{eq:FDash1} can be approximated as follows:
\begin{eqnarray}
F'_1(\rho_{ps})&\approx&-\Gamma (L+1) \left(\frac{\eta\rho_{ps} L \lsp}{\lrd \lsr \psi}+1\right)^{-L}.
\end{eqnarray}
After performing the double derivative with respect to $\rho_{ps}$ and it is given by:
\vspace{-0.15cm}
\begin{eqnarray}
\frac{d^2F'_1(\rho_{ps})}{d\rho_{ps}^2}\approx - \left(\frac{\eta \lsp}{\lrd \lsr \psi}\right)^2 \frac{L^3 \Gamma (L+2)}{\left(\frac{\eta L \lsp \rho_{ps}}{\lrd \lsr \psi}+1\right)^{L+2}}\,.
\end{eqnarray} 
It can be {seen} that $\frac{d^2F'_1(\rho_{ps})}{d\rho_{ps}^2}$ is always negative for all system values.
\subsubsection*{Proof of concavity of function $F_2\left(\rho_{ps}\right)$}
$F_2(\rho_{ps})$ is given by:
\begin{IEEEeqnarray}{rcl}
	F_2(\rho_{ps})&=&{\left(\frac{\lsp \xi}{\lsr \psi}+1\right)^{-L}} \sum_{i=1}^L (-1)^i \binom{L}{i} e^{-\frac{i \lrd \xi}{\rho_{ps}}}.\label{eq:DoubleDerivativeofF1}
\end{IEEEeqnarray}
After applying double derivative w.r.t. $\rho_{ps}$, {we get:}
{\small\begin{IEEEeqnarray}{rcl}
&&\frac{d^2 F_2(\rho_{ps})}{d \rho_{ps}^2}=\sum_{i=1}^L (-1)^i \binom{L}{i}\frac{e^{-\frac{i \lrd (1-\rho_{ps})}{\rho_{ps}}}}{\left(\frac{\lsp (1-\rho_{ps})}{\lsr \psi}+1\right)^L} \bigg(\frac{i^2 \lrd^2}{\rho_{ps}^4}-\frac{2 i \lrd}{\rho_{ps}^3}+\frac{L \lsp (2 i \lrd)}{\rho_{ps}^2 (\lsp (1-\rho_{ps})+\lsr \psi)}+\frac{L (L+1) \lsp^2}{(\lsp (1-\rho_{ps})+\lsr \psi)^2}\bigg)\nonumber.
\end{IEEEeqnarray}}
For smaller values of $\rho_{ps}\rightarrow 0$, terms inside the summation are extremely small because of the term $e^{-\frac{i \lrd (1-\rho_{ps})}{\rho_{ps}}}$. However, for larger values of $\rho_{ps}$ ($\rho_{ps}\rightarrow 1$), terms which contains highest power of $(1-\rho_{ps})$ in the denominator dominate. After retaining only the significant term (i.e. $\frac{L (L+1) \lsp^2}{(\lsp (1-\rho_{ps})+\lsr \psi)^2}$),  and after some manipulations the above equation can be expressed as:
\begin{IEEEeqnarray}{rcl}
&&\frac{d^2F_2(\rho_{ps})}{d \rho_{ps}^2}\approx \sum_{i=1}^L (-1)^i \binom{L}{i}\frac{e^{-\frac{i \lrd (1-\rho_{ps})}{\rho_{ps}}}}{\left(\frac{\lsp (1-\rho_{ps})}{\lsr \psi}+1\right)^L} \left(\frac{L (L+1) \lsp^2}{(\lsp (1-\rho_{ps})+\lsr \psi)^2}\right)\nonumber\\
&=& \frac{\left(1-e^{-\frac{\lrd (1-\rho_{ps})}{\rho_{ps}}}\right)^L-1}{\left(\frac{\lsp (1-\rho_{ps})}{\lsr \psi}+1\right)^{L+2} \frac{\lsr\psi}{\lsp}} \left({L (L+1)}\right).\label{eq:DoubleDerivativeofF2}
\end{IEEEeqnarray}
Right hand side of the above equation can be shown to be always less than $0$, since $\left(1-e^{-{\lrd (1-\rho_{ps})}/{\rho_{ps}}}\right)^L\leq1$. Proof  for the TS-EH case follows in a similar fashion, since $\frac{d^2\tau_{ts}}{d\rho_{ts}^2}$ consists of two function, and both have negative values for $0\leq\rho_{ts}\leq1$.
\section*{Appendix-D:Optimum $\rho$ for L=1}
\setcounter{equation}{0}
\renewcommand{\theequation}{D.\arabic{equation}}
\subsubsection*{For PS-EH}
From \eqref{eq:tauMRCLRPtendInfty}, $\tilde{\tau}$ for PS-EH ($\xi = 1 - \rho_{ps}$,
$\zeta = 1$ and $\beta = \eta \rho_{ps}$) with $L=1$ can be written as:
\begin{IEEEeqnarray}{rcl}
	\tilde{\tau}_{ps}&\overset{L=1}{=}&\frac{R_s}{2}\bigg(1-\bigg(\frac{\lsr e^{-\frac{\lrd (1-\rho_{ps})}{\rho_{ps}}}}{\frac{ \lsp (1-\rho_{ps})}{\psi }+\lsr}+\frac{1}{\left(1+\frac{\lsp}{\lsd\psi }\right)} \frac{\psi \lrd \lsr }{ \lsp \rho_{ps}} e^{\frac{\psi \lrd \lsr }{ \lsp \rho_{ps}}}  E_1\left(\frac{\psi \lrd \lsr }{ \lsp \rho_{ps}}\right)\bigg)\bigg).\nonumber\nonumber
\end{IEEEeqnarray}
In the current form, optimum $\rho_{ps}$ cannot be evaluated analytically. $\tilde{\tau}_{ps}$ can be approximated using the relations
$e^{x}E_1(x)\approx \frac{1}{1+x}$ (for small $x$) \cite[5.1.19]{Abramowitz1964} and $e^{-\frac{\lrd (1-\rho_{ps})}{\rho_{ps}}}\approx \frac{1}{1+\frac{\lrd (1-\rho_{ps})}{\rho_{ps}}}$
to get:
\begin{IEEEeqnarray}{rcl}
\tilde{\tau}_{ps}&\approx& \frac{R_s}{2}\Bigg(1-\frac{1}{\left(1+\frac{\lsp}{\lsd\psi }\right)\left(1+\frac{ \lsp \rho_{ps}}{\psi \lrd \lsr }\right)}-\frac{1}{\left(1+\frac{ \lsp (1-\rho_{ps})}{\psi \lsr}\right)\left(1+\frac{\lrd(1-\rho_{ps})}{\eta\rho_{ps}}\right)}\Bigg).
\end{IEEEeqnarray} 
It is worth noting that the above approximation is valid when ${\lsp} \gg \psi$. 
Finding the optimum $\rho_{ps}$ is still a difficult task since the equation $\frac{d\tilde{\tau}_{ps} }{d\rho_{ps}}=0$ contains $4^{th}$ {power} of $\rho_{ps}$. However, it can be calculated using  standard packages like Mathematica or Matlab. To get a closed form expression, we further approximate the above by utilizing the fact $\frac{\lrd}{\eta\rho_{ps}} \gg 1-\frac{\lrd}{\eta}$ (since $\eta, \rho_{ps}\leq 1$) and  $\frac{ \lsp (1-\rho_{ps})}{\psi \lsr}\gg 1$ {(for typical values of system parameters)} to get:
\begin{IEEEeqnarray}{rcl}
	\tilde{\tau}_{ps}&\approx&\frac{R_s}{2}\Bigg( 1-\frac{1}{\left(1+\frac{\lsp}{\lsd\psi }\right)\left(1+\frac{ \lsp \rho_{ps}}{\psi \lrd \lsr }\right)}-\frac{1}{\frac{ \lsp (1-\rho_{ps})}{\psi \lsr}\frac{\lrd}{\eta\rho_{ps}}}\Bigg).\label{eq:TauMRCApprox}
\end{IEEEeqnarray} 
Now, we can obtained $\rho^*_{ps}$ by solving $\frac{d	\tilde{\tau}_{ps}}{d\rho_{ps}}=0$ to get \eqref{eq:rhoPS}.
\subsubsection*{For TS-EH}
Using the relation $e^{x}E_1(x)\approx \frac{1}{1+x}$, $\tilde{\tau}$ for TS-EH ($\xi = 1$,
$\zeta = 1-\rho_{ts}$ and $\beta = 2\eta \rho_{ts}/(1-\rho_{ts})$) with $L=1$,  \eqref{eq:tauMRCLRPtendInfty} can be approximated as:
\begin{IEEEeqnarray}{rcl}
\tilde{\tau}_{ts}&\approx &\frac{(1-\rho_{ts}) \text{Rs}}{2}  \bigg(\hspace{-0.2cm}-\frac{\psi \lsr  e^{-\frac{\lrd (1-\rho_{ts})}{2 \eta \rho_{ts}}}}{ \lsp}-\frac{1}{\left(\frac{ \lsp}{\psi \lsd }+1\right) \left(\frac{2 \eta  \lsp \rho_{ts}}{\psi \lrd \lsr  (1-\rho_{ts})}+1\right)}+1\bigg).
\end{IEEEeqnarray}
It can be further approximated since $\left({\left(\frac{ \lsp}{\psi \lsd }+1\right) \left(\frac{2 \eta  \lsp \rho_{ts}}{\psi \lrd \lsr  (1-\rho_{ts})}+1\right)}\right)^{-1}$ $\gg$ $ \frac{\psi \lsr  }{ \lsp}e^{-\frac{\lrd (1-\rho_{ts})}{2 \eta \rho_{ts}}}$ to get:
\begin{equation}
\tilde{\tau}_{ts}\approx \frac{(1-\rho_{ts})R_s}{2}\left(\hspace{-0.1cm}1-\frac{1/\left(1+{ \lsp}/({\psi \lsd })\right)}{ \left(\frac{2 \eta  \lsp \rho_{ts}}{\psi \lrd \lsr  (1-\rho_{ts})}+1\right)}\hspace{-0.1cm}\right).\label{eq:tau_tsApprox}
\end{equation}
Solving $\frac{d\tilde{\tau}_{ts}}{d\rho_{ts}}=0$ results in \eqref{eq:rhoTS}.
\section*{Appendix-E: Proof of \eqref{eq:SumSum_ijxExpijxE1ijx}}
\setcounter{equation}{0}
\renewcommand{\theequation}{E.\arabic{equation}}
In \eqref{eq:SumSum_ijxExpijxE1ijx}, let
$\sum _{i=1}^L (-1)^{i+1} \binom{L}{i}L \,e^{\frac{\psi i \lrd \lsr }{\beta  \lsp}} E_{L+1}\left(\frac{\psi i \lrd \lsr}{\beta  \lsp}\right)$, and for higher $L$ and small argument (i.e. $x$) $e^x E_L(x) \approx 1/(x+L-1)$   \cite[5.1.4]{Abramowitz1964}, the above equation can be rewritten as:
\begin{IEEEeqnarray}{rcl}
\sum _{i=1}^L (-1)^{i+1} \binom{L}{i}L \,e^{\frac{\psi i \lrd \lsr }{\beta  \lsp}} E_{L+1}\left(\frac{\psi i \lrd \lsr}{\beta  \lsp}\right)&\approx& \sum _{i=1}^L (-1)^{i+1} \binom{L}{i}\frac{L}{\frac{\psi i \lrd \lsr }{\beta  \lsp}+L}\label{eq:Temp1}\\
&&=1-\frac{\Gamma (L+1) \Gamma \left(\frac{\beta L \lsp}{\lrd \lsr \psi}+1\right)}{\Gamma \left(\frac{\beta \lsp L}{\lrd \lsr \psi}+L+1\right)}. \label{eq:Temp2}
\end{IEEEeqnarray}
We get \eqref{eq:Temp2} from \eqref{eq:Temp1} by applying the relation \eqref{eq:SumRel}. By using relation \eqref{eq:RatoOfGamma}, right hand side (R.H.S) of the above equation can be approximated as $ 1-\Gamma(L)\left(\frac{\lrd \lsr \psi}{\beta L \lsp}\right)^L$.
For the general parameter values $\Gamma(L)\left(\frac{\lrd \lsr \psi}{\beta \lsp}\right)^L\ll L^L$. $\Gamma(L)\left(\frac{\lrd \lsr \psi}{\beta L \lsp}\right)^L\rightarrow 0$. The value of $\sum _{i=1}^L (-1)^{i+1}$ $ \binom{L}{i}L \,e^{\frac{\psi i \lrd \lsr }{\beta  \lsp}} E_{L+1}\left(\frac{\psi i \lrd \lsr}{\beta  \lsp}\right)\rightarrow 1$.
\section*{Appendix-F:Proof of Lemma \ref{Lem:TauINC_Concave}}
\setcounter{equation}{0}
\renewcommand{\theequation}{F.\arabic{equation}}
For PS-EH case (i.e. $\zeta=1$ and $\xi=1-\rho_{ps}$) (from \eqref{eq:tauIn_2}) it is clear that $\tau_{in}$ is given approximately by:
{\begin{IEEEeqnarray}{rcl}
	\tau_{in-ps}&\approx&\tau+\frac{0.5 R_s}{1+\frac{\lsd \psi }{\lsp }}\bigg(1+\left(1+\frac{\lsd(1-\rho_{ps})}{\lsr}+\frac{\lsp(1-\rho_{ps})}{\lsr\psi}\right)^{-L}\bigg).\nonumber
\end{IEEEeqnarray} }
{Substituting} the value of $\tau$ (i.e. $\tilde{\tau}$ for PS-EH case) from \eqref{eq:tauMRCLRPtendInfty}, $\tau_{in-ps} $ is rewritten as follows:
		\begin{IEEEeqnarray}{rcl}
		\tau_{in}&\approx&\frac{R_s\zeta}{2}\Bigg(1+\left(1+\frac{\lsp}{\lsd \psi}\right)^{-1}\underbrace{\Big(\sum _{i=1}^L (-1)^{i+1}\binom{L}{i} L e^{\frac{\psi i \lrd \lsr }{\beta \lsp}} E_{L+1}\Big(\frac{\psi i \lrd \lsr }{\beta \lsp}\Big)-1\Big)}_{F_1(\rho_{ps})}		\vspace{-0.35cm}\nonumber\\
		&&+\underbrace{\overbrace{\frac{\big(1-e^{-\frac{\lrd \xi}{\beta}}\big)^L-1}{\left(\frac{\lsp \xi}{\lsr \psi}+1\right)^L}}^{F_2(\rho_{ps})}+\frac{1}{1+\frac{\lsd \psi }{\lsp }}\left(1+\left(1+\frac{\lsd\xi}{\lsr}+\frac{\lsp\xi}{\lsr\psi}\right)^{-L}\right)}_{F_3(\rho_{ps})}\Bigg).\label{eq:tauIn_3}
	\end{IEEEeqnarray} 
For concavity, we need to show that $\frac{d^2\tau_{in-ps}}{d\rho_{ps}^2}\leq 0$. $\frac{d^2\tau_{in-ps}}{d\rho_{ps}^2}$ is given by:
\begin{eqnarray}
\frac{d^2\tau_{in-ps}}{d\rho_{ps}^2} = \frac{R_s}{2}\left(\frac{d^2F_1(\rho_{ps})}{d\rho_{ps}^2} +\frac{d^2F_3(\rho_{ps})}{d\rho_{ps}^2}\right).\label{eq:DoubleDerivativeofTauIn}
\end{eqnarray}
Using the derived expressions for $\frac{d^2F_2(\rho_{ps})}{d\rho_{ps}^2}$ in Appendix-C from \eqref{eq:DoubleDerivativeofF2}, we can write $\frac{d^2 F_3(\rho_{ps})}{d\rho_{ps}^2}$ as follows:
\begin{equation}
\frac{d^2F_3(\rho_{ps})}{d\rho_{ps}^2}\approx\frac{(L+1) L  \left(\left(1-e^{-{\lrd (1-\rho_{ps})}/{\rho_{ps}}}\right)^L-1\right)}{\left({\lsr \psi}/{\lsp}\right)^2\left({\lsp (1-\rho_{ps})}/({\lsr \psi})+1\right)^{L+2}}+\frac{L (L+1) \left(\frac{\lsd}{\lsr}+\frac{\lsp}{\lsr \psi}\right)^2 }{\left(\frac{\lsd \psi}{\lsp}+1\right)\left(\frac{\lsd (1-\rho_{ps})}{\lsr}+\frac{\lsp (1-\rho_{ps})}{\lsr \psi}+1\right)^{L+2}}.\label{eq:ddF3}
\end{equation}
{For} (${\lsp}\gg{\psi}$) and for ($0<\rho_{ps}<1$), above expression can be approximated as:
\begin{IEEEeqnarray}{rcl}
	\frac{d^2F_3(\rho_{ps})}{d\rho_{ps}^2}&\approx&\frac{(L+1) L \left(\left(1-e^{-{\lrd (1-\rho_{ps})}/{\rho_{ps}}}\right)^L-1\right)}{(1-\rho_{ps})^{L+2}\left({\lsp }/({\lsr \psi})\right)^{L}}+\frac{L (L+1) }{(1-\rho)^{L+2}\left(\frac{\lsd \psi}{\lsp}+1\right)\left(\frac{\lsd }{\lsr}+\frac{\lsp }{\lsr \psi}\right)^{L}}.\nonumber
\end{IEEEeqnarray}
Approximation of the above equation from \eqref{eq:ddF3}, holds unless $\rho_{ps}=1$ or close to 1. However, {the} network never operates {in this region} as in this case R will not be able to decode signals from S.
It can be observed that, the above equation has negative values for $\rho_{ps}$ {so} that:
\begin{eqnarray}
\frac{d^2F_3(\rho_{ps})}{d\rho_{ps}^2}\leq 0.\label{eq:ZeroCrossing}
\end{eqnarray} 
By solving the above equation, range of $\rho_{ps}$ is given as:
{\begin{eqnarray}
\rho_{ps} \geq \rho_{zc}=\Bigg(\hspace{-0.1cm}{1-\frac{1}{\lrd}\log \Bigg(\hspace{-0.1cm}1-\Bigg(\hspace{-0.1cm}1-\bigg(\frac{\lsd\psi}{\lsp}+1\bigg)^{-L-1}\Bigg)^{1/L}\Bigg)}\hspace{-0.1cm}\Bigg)^{-1}\hspace{-0.2cm}, \nonumber
\end{eqnarray}}
where $\rho_{zc}$ is the solution of \eqref{eq:ZeroCrossing} with equality. The subscript $zc$ highlights the zero crossing point of \eqref{eq:ZeroCrossing}. We note that for $\rho$<$\rho_{zc}$, $\frac{d^2F_3(\rho_{ps})}{d\rho_{ps}^2}$ have finite positive values. However, they are extremely small (in the order of $10^{-2}$) for the general system parameters which is  $\psi$$\ll$$\lsp$ and $\lsp$$\gg$$\lsd$, $\frac{d^2F_3(\rho_{ps})}{d\rho_{ps}^2}$.
From \eqref{eq:DoubleDerivativeofF1}, we recall that $\frac{dF_1(\rho_{ps})}{d\rho_{ps}^2}\rightarrow -\infty$ for $\rho_{ps}\rightarrow 0$ $\implies \frac{d^2\tau_{in-ps}}{d\rho_{ps}^2}<0$. Since $\frac{d^2\tau_{in-ps}}{d\rho_{ps}^2}$ is the summation of the $\frac{d^2F_1(\rho_{ps})}{d\rho_{ps}^2}$ and $\frac{d^2F_3(\rho_{ps})}{d\rho_{ps}^2}$, and since $\frac{d^2F_1(\rho_{ps})}{d\rho_{ps}^2}\leq \frac{d^2F_3(\rho_{ps})}{d\rho_{ps}^2}$ for the range $0\leq\rho_{ps}\leq\rho_{zc}$ (from the observation), it can be inferred that $\frac{d^2\tau_{in-ps}}{d\rho_{ps}^2}<0$ \, $\forall\, \rho_{ps}$.
\vspace{-0.1cm}
\section*{Appendix-G: Proof of Lemma \ref{Lem:rho_ts_in}}
	For the PS-EH case, using \eqref{eq:tauIn_2} and \eqref{eq:TauMRCApprox}, the approximated $\tau_{in} $ is given by:
\begin{equation}
	\tau_{in-ps}\approx \frac{R_s}{2}\Bigg(1-\Bigg(\hspace{-0.1cm}\frac{1}{\left(1+\frac{\lsp}{\lsd\psi}\right)\left(1+\frac{\eta  \lsp \rho_{ps}}{\psi \lrd \lsr }\right)}+\frac{1}{\frac{\lrd}{\eta \rho_{ps}}\frac{ \lsp (1-\rho_{ps})}{\psi \lsr }}+\frac{1}{\frac{\psi}{ \lsp} \left(\lsd+\frac{\lsr}{1-\rho_{ps}}\right)+1}\Bigg)\Bigg)+R_s q_2,\label{eq:TauInPSApproxApp1}
\end{equation}
where $q_2$ is as defined in \eqref{eq:tauIn}. 	An optimum value of $\rho_{ps}$ can be determined numerically. However, an approximated closed-form expression can be derived  {for the condition $\psi \ll \lsp$}  since $\frac{\psi}{ \lsp} \left(\lsd+\frac{\lsr}{1-\rho_{ps}}\right)\ll1$ (for $\rho_{ps}<1$). \eqref{eq:TauInPSApproxApp1} is now expressed as:
{\begin{IEEEeqnarray}{rcl}
\tau_{in-ps} \approx R_s q_2+\underbrace{\frac{R_s}{2}\Bigg(1-\Bigg(\frac{1}{\left(1+\frac{\lsp}{\lsd\psi}\right)\left(1+\frac{\eta  \lsp \rho_{ps}}{\psi \lrd \lsr }\right)}+\frac{1}{\frac{\lrd}{\eta \rho_{ps}}\frac{ \lsp (1-\rho_{ps})}{\psi \lsr }}\Bigg)\Bigg)}_{\tau_{ps}}.\nonumber
\end{IEEEeqnarray}}
We verify later through simulations that this approximate value is very close to the optimum $\rho_{in-ps}^{*}$. Again from \eqref{eq:TauMRCApprox}, $\tau_{in-ps}$ can be rewritten as: $\tau_{in-ps}=\tau_{ps}+R_s q_2$, since $q_2$ is not a function of $\rho_{ps}$. Clearly, $\rho_{in-ps} = \rho_{ps}$ since:
\begin{equation}
\frac{d\tau_{in-ps}}{d\rho_{ps}} \approx \frac{d\tau_{ps}}{d\rho_{ps}}=0.
\end{equation}
For TS-EH case with $\lsr \gg \lsd$, $\tau_{in}$ is approximated as:
\begin{IEEEeqnarray}{rcl}
	\tau_{in-ts} &\approx& \frac{(1-\rho_{ps})R_s}{2}\Bigg(1-\frac{1}{\left(\frac{ \lsp}{\psi \lsd }+1\right) \left(\frac{2 \eta  \lsp \rho_{ps}}{\psi \lrd \lsr  (1-\rho_{ps})}+1\right)}-\frac{1}{\frac{\psi\lsd}{ \lsp}+1}\Bigg)+R_s q_2\,.\label{eq:TauInTSApproxApp1}
\end{IEEEeqnarray}
From \eqref{eq:tau_tsApprox}, the above expression is written as: $\tau_{in-ts}=\tau_{ts}+\frac{1-\rho_{ps}}{2}R_s\,q_2$. We first note that $q_2$ is not a function of $\rho_{ps}$. Now the optimum value of $\rho_{ps}$ in \eqref{eq:rho_ts_in} is the solution to the following equation:
\begin{equation}
\frac{d\tau_{in-ts}}{d\rho_{ps}} \approx \frac{d\tau_{cc-ts}}{d\rho_{ps}}-\frac{0.5 R_s}{1+\frac{\psi\lsd }{\lsp }}=0.
\end{equation}	
\vspace{-0.4cm}
\bibliographystyle{ieeetr}
\bibliography{WPMCExt}
\end{document}